\documentclass{article}

\usepackage[english]{babel}
\usepackage[letterpaper,top=2cm,bottom=2cm,left=3cm,right=3cm,marginparwidth=1.75cm]{geometry}
\usepackage{amsmath}
\usepackage{amssymb}
\usepackage{amsthm}
\usepackage{graphicx}
\usepackage[colorlinks=true, allcolors=blue]{hyperref}

\usepackage{biblatex}
\usepackage[utf8]{inputenc}
\usepackage[english]{babel}
\usepackage[ruled]{algorithm2e}
\usepackage{xcolor}
\usepackage{setspace}
\usepackage{apalike}
\usepackage{caption}
\usepackage{soul}
\usepackage{changes}

\usepackage{float}
\usepackage{tabularx}
\usepackage{multirow}
\newcolumntype{Y}{>{\centering\arraybackslash}X}

\usepackage{subcaption}
\usepackage{array}
\makeatother
\newcommand{\half}{\frac{1}{2}}
\newcommand{\cu}{\textbf}
\newcommand{\cuu}{\boldsymbol}
\newcommand{\xie}{\textit}

\newcommand*{\rom}[1]{#1}

\newtheorem {theorem} {\bf Theorem}
\newcommand{\Exp}{\text{Exp}}
\newcommand{\Log}{\text{Log}}

\newcommand{\M}{\mathcal{M}}
\newcommand{\N}{\mathcal{N}}
\newcommand{\A}{\mathcal{A}}

\newcommand{\T}{\mathcal{T}}

\newcommand{\X}{\mathcal{X}}
\newcommand{\D}{\mathcal{D}}

\newcommand{\R}{\mathbb{R}}
\newcommand{\Hilbert}{\mathbb{H}}
\newcommand{\rma}{Riemannian manifold}
\newcommand{\inj}[1]{\textrm{inj}_{#1}}

\newcommand{\vesub}[2]{\mbox{{\boldmath ${#1}$}$_{#2}$}}

\definecolor{BrickRed}{cmyk}{0, .89, .94, .28}
\definecolor{DarkBlue}{rgb}{0.0, 0.08, 0.45}

\title{Wrapped Gaussian Process Functional Regression Model for Batch Data on Riemannian Manifolds}
\author{Jinzhao Liu, Chao Liu, Jian Qing Shi$^*$, Tom Nye}
\date{\vspace{-5ex}}
\begin{document}

\begin{sloppypar}

\maketitle

\begin{abstract}
Regression is an essential and fundamental methodology in statistical analysis. The majority of the literature focuses on linear and nonlinear regression in the context of the Euclidean space. However, regression models in non-Euclidean spaces deserve more attention due to collection of increasing volumes of manifold-valued data. In this context, this paper proposes a concurrent functional regression model for batch data on \rma s by estimating both mean structure and covariance structure simultaneously. The response variable is assumed to follow a wrapped Gaussian process distribution. Nonlinear relationships between manifold-valued response variables and multiple Euclidean covariates can be captured by this model in which the covariates can be functional and/or scalar. The performance of our model has been tested on both simulated data and real data, showing it is an effective and efficient tool in conducting functional data regression on Riemannian manifolds.
\end{abstract}
 
\section{Introduction}
\begin{figure}
    
    \centering

    \includegraphics[width=1\textwidth,trim={5cm 0cm 0cm 5cm},clip]{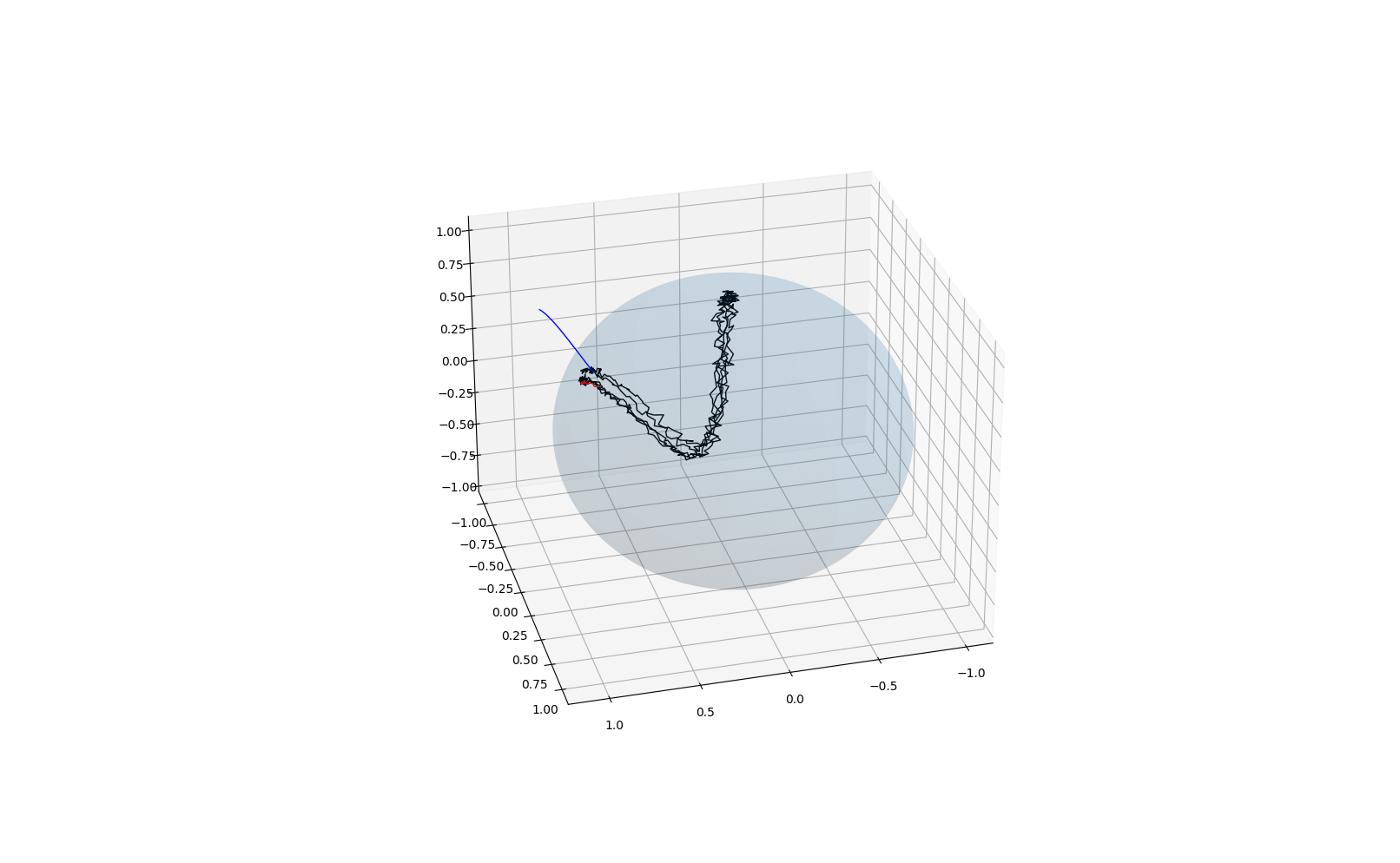}

    \caption{The black curves represent manifold-valued data, while the red and blue curves depict the predictions of Gaussian process regressions with and without a manifold structure, respectively. This illustrates the impact of incorporating a manifold structure into the regression model, as evidenced by the alignment of the red curve with the manifold-valued data, in contrast to the deviation of the blue curve. This underscores the importance of considering manifold structures in Gaussian process regression models for accurate prediction and data representation.}
    
    \label{fig:diff_models_on_sphere}
    
\end{figure}

Regression models are ubiquitous and powerful tools in data analysis to reveal the relationship between independent variables and dependent variables. Most well known regression models are formulated under the assumption that the variables all lie in Euclidean vector spaces. However, statistical models of manifold-valued data are gaining popularity in various fields of scientific analysis, such as computational social science, medical imaging analysis and computer vision \cite{Bronstein2017GeometricDL}. Together with the significant increase in manifold-valued data observed by current instrumentation, generalizations of Euclidean regression models to manifold-valued settings are needed. If the manifold structure is not accounted for then the reliability of model predictions in the target spaces is compromised. Consider, for instance, Figure \ref{fig:diff_models_on_sphere}, where a regression model is trained on manifold-valued data (represented by the black curves on a S2). The predictions of this model (illustrated by the blue curve) deviate from the manifold. However, when the manifold structure is incorporated into the model, the predictions (depicted by the red curve) align with the target manifold. This underscores the necessity of employing non-Euclidean structure for modelling manifold-valued data. An additional example demonstrating this principle in Kendall's shape space is provided in the Appendix.

In recent years, functional data have garnered increasing attention due to the continuous or intermittent recording of data at discrete time points. These data often exhibit non-linearity, and a common assumption is that they lie on a nonlinear manifold. For instance, image data, which can be influenced by random domain shifts, are known to reside in such manifolds . However, analyzing functional data poses challenges, particularly when dealing with spaces lacking global or local linear structures. Traditional methods like functional principal component analysis become infeasible in these cases. One specific difficulty arises from defining a suitable metric for data residing on a Riemannian manifold. Researchers have explored alternative approaches to address this issue. For example, \cite{dubey2021modeling} propose using point-wise Fr\'{e}chet means for non-Euclidean time-varying random objects in general metric spaces. Additionally, they introduce point-wise distance trajectories between individual time courses and the estimated Fr\'{e}chet mean trajectory. This innovative approach allows for a representation of time-varying random objects using functional data. In summary, the study of functional data on nonlinear manifolds presents exciting opportunities for advancing statistical methodologies and understanding complex data structures.

According to the manifold hypothesis, functional data can be mapped onto a low-dimensional nonlinear manifold, which is also known as manifold learning \cite{roweis2000nonlinear}. 
Gaussian processes (GPs) have proven to be powerful tools for modeling complex data structures. While GPs are commonly applied in Euclidean spaces, recent research has extended their applicability to Riemannian manifolds. \cite{borovitskiy2020matern} utilize the spectral theory of the Laplace-Beltrami operator to compute Mat\'{e}rn kernels on Riemannian manifolds. These kernels, widely used in physical sciences, capture both smoothness and spatial correlations. The spectral approach allows for efficient computation of Mat\'{e}rn covariance functions, enabling GPs to model data residing on curved surfaces. \cite{hutchinson2021vector} introduce a vector-valued GP equipped with a matrix-valued kernel. This novel framework enables modeling of vector fields on Riemannian manifolds. Applications include geodesic flow modeling, fluid dynamics, and other scenarios where vector-valued data arise. Recently, \cite{azangulov2022stationary} and \cite{azangulov2023stationary} extend stationary GPs to compact Lie groups and non-compact symmetric spaces. These spaces play a critical role in spatiotemporal modeling. Those models define a metric based on geodesic distances between predictors, accommodating both functional and scalar predictors. In addition, \cite{lin2021functional} propose a regression model for functional data on manifolds. While existing approaches assume Euclidean response variables, certain problems involve responses residing directly on Riemannian manifolds. For instance, predicting flight trajectories based on flight time. In contrast to previous models, we propose a regression framework that maps Euclidean-valued predictors to manifold-valued functional responses.

A number of statistical approaches to analysis of manifold-valued data exist in the literature, especially in the context of principal component analysis (PCA). For example, \cite{fletcher2004principal} introduced principal geodesic analysis which is a generalisation of PCA on \rma s. It replaces the first principal component with a principal geodesic constrained to pass through the intrinsic mean, with calculations performed approximately in Euclidean tangent space. \cite{lila2016smooth} proposed a PCA technique for functional data on $2$-dimensional Riemannian manifolds in which the authors adopt a regularization method by a smoothing penalty coherent with geodesic distances. Subsequently, \cite{dai2018principal} formulated Riemannian functional principal component analysis by mapping data to tangent spaces via the Riemannian logarithm map and then performing multivariate functional principal component analysis within the tangent space. 

In addition to generalizations of PCA to manifold data, generalizations of linear regression have also been studied. Similar to manifold-based PCA, the regression line is typically replaced with a geodesic within the manifold, which gives rise to more challenging optimisation problems in order to fit the models than in the standard Euclidean setting. Geodesic regression \cite{fletcher2013geodesic} is a generalization of linear regression to a manifold-valued setting in which an univariate independent variable is considered in $\mathbb{R}$ and dependent variables lie in a \rma. The author derived a gradient descent algorithm for model fitting via derivatives of the exponential map and Jacobi fields, because the least square method has no analytical solution under this setting. \cite{kim2014multivariate} then extended geodesic regression to the multivariate setting in which the independent variables are in $\mathbb{R}^n$ and the dependent variable is still manifold-valued. In addition, the authors proposed a variational gradient descent method based on parallel transport which is more convenient for high-dimensional independent variables. In the context of shape analysis, \cite{nava2020geodesic} provided an analytic approach to geodesic regression in Kendall's shape space. Multiple linear regression has also been generalised by \cite{petersen2019frechet} for complex random objects in metric spaces with independent variables in $\mathbb{R}^n$. Using a least-squares approach, the authors derived asymptotic rates of convergence.

However, in many applications, non-Euclidean data cannot always be assumed to follow form of linear relationship. \cite{Banerjee_2016_CVPR} introduced a kernel-based nonlinear regression model for both manifold-valued independent variables and manifold-valued dependent variables. \cite{hinkle2012polynomial} extended polynomial regression to \rma s by introducing a class of curves on \rma s which generalize geodesics and are analogs of polynomial curves in Euclidean space. \cite{cornea2017regression} developed a regression model for a manifold-valued response variable in Riemannian symmetric spaces and covariates in Euclidean spaces with applications in medical imaging analysis. Moreover, \cite{pigoli2016kriging} introduced an additive linear model for manifold-valued data which is based on the exponential map. In particular, they transform the manifold-valued data onto tangent spaces and then estimate the parameters of the additive linear model by a generalised least squares method.

Gaussian process regression (GPR) is a powerful non-linear and non-parametric model widely used for learning probability distributions over unknown functions. Operating within a Bayesian framework, GPR assumes that both the prior and likelihood follow normal distributions, allowing for the derivation of a posterior distribution that is also Gaussian. 
This ash been extended to solve the problems involving non-Gaussian data \cite{wang2014generalized} and the use of other prior processes \cite{wang2021general}. In addition, researchers have extended GPR to accommodate multi-output scenarios. Notable contributions include the work of \cite{boyle2005dependent}, \cite{van2020framework}, and \cite{alvarez2012kernels}, who have explored multi-output Gaussian process regression. Over the past decade, substantial progress has been made in developing Gaussian process regression models specifically tailored for manifolds, thereby expanding the model’s scope and applicability. One intriguing approach is the wrapped Gaussian process regression proposed by \cite{mallasto2018wrapped}. In this framework, the Riemannian manifold is linearized via the logrithm map, projecting data points onto a tangent space. This results in a non-parametric regression model with a probabilistic foundation. Importantly, the computational cost of wrapped Gaussian process regression remains relatively low, as it involves additional calculations only for the exponential map and logarithm map of each manifold-valued data point. The key innovation lies in defining a Gaussian distribution directly on the Riemannian manifold, leveraging insights from real-valued multivariate Gaussian distributions and the exponential map. This novel approach opens up avenues for Gaussian process regression on Riemannian manifolds. Within the Bayesian framework, researchers have derived manifold-valued posterior distributions, albeit under certain assumptions (e.g., the requirement of infinite injectivity radius). 

To the best of our knowledge, there is little current literature about regression models for functional batch data on \rma s\ within a probabilistic framework. In light of this, we provide a nonlinear non-parametric regression model with uncertainty for batch data on a smooth \rma. Called the \emph{wrapped Gaussian process functional regression model} (WGPFR), it models the relationship of a functional manifold-valued response and mixed (scalar and functional) Euclidean-valued predictors by representing the mean structure and the covariance structure as different model terms. Specifically, a functional regression model is used for the mean structure with a functional manifold-valued response and scalar covariates, while a wrapped Gaussian process has been used for the covariance structure with functional covariates based on \cite{mallasto2018wrapped} who extend Gaussian process regression on \rma s. In this way, the proposed WGPFR model extends the research field of non-parametric regression on manifolds.

The rest of this paper is organized as follows. In section \ref{background}, some basic concepts of functional data, \rma s, Gaussian process regression and wrapped Gaussian process regression are reviewed. In section \ref{sec:model}, we propose our model together with a method for inference via an efficient iterative algorithm.  Numerical experiments and real data analysis are reported in Section \ref{sec:numerical}. Finally, we draw conclusions and discuss the results in Section \ref{sec:discussion}.

\section{Background}
\label{background}

\subsection{Functional Data}

Functional data analysis is a field of study wherein each observation is a random function, manifesting as a curve, surface, or other variations over a continuum. This analysis is particularly pertinent to data recorded over identical intervals, with consistent frequency and numerous repeated observations. While data are frequently modelled by parametric models incorporating randomness in the realms of statistics or machine learning, functional data analysis contemplates a smooth process observed at discrete time points, offering greater flexibility and reduced parametric constraints, see e.g. \cite{ramsay2004functional} and \cite{yao2005functional}.

Furthermore, functional data can be procured from a diverse range of domains. For instance, in finance, volatility surfaces serve as a source of functional data. In the field of biomedical research, time-varying physiological signals yield functional data. Functional Magnetic Resonance Imaging (FMRI) is yet another domain that generates functional data. These examples underscore the broad applicability and interdisciplinary nature of functional data analysis.

\subsection{Preliminaries for Riemannian Manifolds}

In this section, we review a few basic concepts of \rma s, wrapped Gaussian distributions, wrapped Gaussian processes on manifolds and then set up basic notations. More detail on basic Riemannian geometry can be found in standard texts, such as that by \cite{do1992mathematics}. 

\subsubsection{Concepts and Notation} 
\label{sec:notation}

Given a $d$-dimensional smooth differentiable manifold $\M$ and a tangent space $T_p\M$, for $p\in\M$, a \xie{Riemannian metric} $g_p:T_p\M\times T_p\M \rightarrow \mathbb{R}$ on $\M$ is a family of positive definite inner products which vary smoothly with $p\in\M$. Equipped with this Riemannian metric, we call the pair $(\M, g)$ a \emph{Riemannian manifold}.

The \xie{tangent bundle}  of  $\M$ is defined as a set  $\mathcal{TM}=\cup_{p\in \M} (p\times T_p\M)$, where $p$ is a point in $\M$ and $v\in T_p\M$ is a tangent vector at $p$. If $\gamma$ is a smooth curve in $\mathcal{M}$, then its length is defined as the integral of $\|\partial \gamma/\partial t\|$ where the norm is computed via the Riemannian metric at $\gamma(t)$. For any pair $(p,v)\in \mathcal{TM}$ with $\|v\|$ sufficiently small, there is a unique curve $\gamma:[0,1]\rightarrow \mathcal{M}$ of minimal length between two points  $p\in\M$ and $q\in\M$, with initial conditions  $\gamma(0)=p$, $\gamma'(0)=v$ and $\gamma(1)=q$. Such a curve is called a \xie{geodesic}. The  \xie{exponential map} $\Exp(p,v):\M\times T_p\M \rightarrow \M$ at $p\in \M$ is defined as $\Exp(p, v)=\gamma(1)$ for sufficiently small $v\in T_p\M$. 
Consider the ball of largest radius around the origin in $T_p\M$ on which  $\Exp(p, v)$ is defined and let $V(p)\subset\M$ denote the image of $\Exp(p,\cdot)$ on this ball. 
Then the exponential map has an inverse on $V(p)$, called the  \xie{Riemannian logarithm map} at $p$, $\Log(p,\cdot): V(p) \rightarrow T_p\M$. For any point $q\in V(p)$, the \xie{geodesic distance} of $p\in\M$ and $q\in\M$ is given by $d_{\mathcal{M}}(p,q)=\|\Log(p,q)\|$. 

\begin{figure}[h]
    \centering
    \includegraphics[width=150mm,scale=0.5]{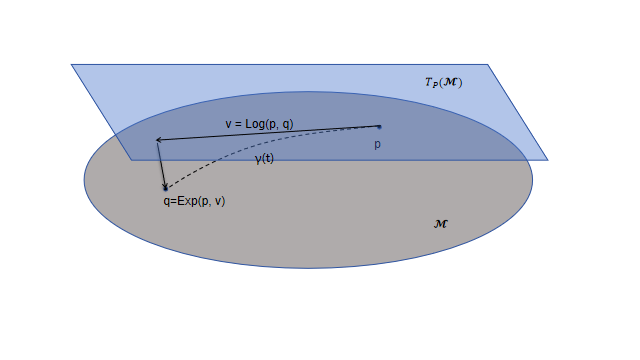}
    \caption{Exponential map and Logarithm map on $S^2$.}
\label{fig:exp_log_map}
\end{figure}

Figure \ref{fig:exp_log_map} shows the concepts above. For example, $p$ and $q$ are two points on a \rma \ $\mathcal{M}$, the tangent space at $p$ is denoted as $T_p(\mathcal{M})$ and $\gamma(t)$ refers to a geodesic between $p$ and $q$. The tangent vector $v$ is from $p$ to $q$ at $T_p(\mathcal{M})$ and $p$ moves towarding $v$ with distance $\|v\|$ is $q$, which represents $\Log(p, q)$ and $\Exp(p, v)$ respectively.

The exponential map and its inverse are analogs of vector space operations in the following way. Suppose $p_{\M}$ and $q_{\M}$ refer to two different points on $\M$ linked by a geodesic $\gamma$, and suppose the tangent vector from the former to the latter is denoted by $pq_{\M}=(\frac{d\gamma}{dt})_p$. In addition, suppose $p_E$ and $q_E$ refer to two Euclidean data points and the vector from the former to the latter is given by $pq_E$. Table \ref{tab:operators_between_euclidean_manifold} shows the relationship of operations between \rma s and Euclidean spaces.
 
\begin{table}[h]
 \centering
 	\caption{Analogs between basic operation in Euclidean space and \rma.}
 	\label{tab:operators_between_euclidean_manifold}
 	\begin{tabularx}{\textwidth}{Y Y Y}
 		\hline
 		&Euclidean Space &Riemannian Manifold\\
 		\hline
 		Addition  &$q_E=p_E+v_{pq}$ &$q_\M=\Exp(p_\M,v_{pq})$  \\
 		Subtraction   &$v_{pq}=p_E-q_E$ &$v_{pq}=\Log(p_\M,q_\M)$ \\
 		Distance  &$\Vert p_E-q_E\Vert$ &$\Vert\Log(p_\M,q_\M)\Vert$
 		\\
 		\hline
 	\end{tabularx}
\end{table}

If $(\M, d_{\M})$ is a \xie{complete metric space} that every Cauchy sequence of points in $\M$ has a limit that is also in $\M$, then the Hopf-Rinow theorem shows that every pair of points $p,q\in\mathcal{M}$ is joined by at least one geodesic, a condition known as \xie{geodesic completeness}, and, equivalently, $\Exp(p,v)$ is defined for all $(p,v)\in\mathcal{TM}$. However, given two points on $\mathcal{M}$, the geodesic between the points may not be unique, even if $\mathcal{M}$ is geodesically complete. 
The \xie{injectivity radius} at $p\in\mathcal{M}$, denoted $\inj{p}$, is defined to be the radius of the largest ball at the origin of $T_p\mathcal{M}$ on which $\Exp(p,\cdot)$ is a diffeomorphism. It follows that the geodesic ball at $p$ of radius $\inj{p}$ is contained within $V(p)$ under which the Wrapped Gaussian process regression can be derived to a analytical form. In the following, we assume the data are on a Riemannian manifold with infinite injectivity radius, such as spheres and Kendall's shape spaces.

Some examples of \rma s are shown in Appendix, including explicit formulae for the exponential and logarithm maps in certain cases.  

\subsection{Functional Data on Riemannian Manifolds}

To study functional data located on \rma s, \cite{dai2018principal} considered a $d$-dimensional complete Riemannian submanifold $\M^\#$ of a Euclidean space $\R^{d_0}$, with a geodesic distance $d_{\M}$ and a probability space $(\Omega, \A, P)$ with sample space $\Omega$, $\sigma$-algebra $\A$ and probability $P$. The sample space of all \xie{$\M$-valued continuous functions} on a compact interval $\T \subset \R$ is denoted by $\X=\{ y: \T \rightarrow \M\ |\ y \in \mathcal{C}(\T)\}$. Thus, we define {\it $\M$-valued random functions} to be functions $y(t, \omega)$, $y: \T \times \Omega \rightarrow \M$, satisfying $y(\cdot, \omega) \in \X$. Let $\Hilbert=\{v: \T \rightarrow \ \mathbb{R}^{d}\ | \ \int_{\T}v(t)^Tv(t)dt < \infty\}$ refers to an ambient $L^2$ Hilbert space of $d$-dimensional square integrable functions on $\mathcal{T}$. The inner product is defined as $\langle v,u\rangle=\int_{\T}v(t)^Tu(t)dt$ and the norm is given by $\|v\|=\langle v,v\rangle^{\half}$ for $v,u\in \Hilbert$.

\subsection{Gaussian Process Regression and Wrapped Gaussian Process Regression}

\subsubsection{Euclidean Gaussian Process Regression}

As described in \cite{rasmussen2003gaussian}, a general Gaussian process regression model is defined as 
\begin{gather}
y=f(\cuu{x})+\epsilon
\end{gather}
 where  $\cuu{x}\in \mathbb{R}^Q$, $y \in \mathbb{R}$, $\epsilon\sim \N(0,\sigma^2)$.
  The regression function $f$ is assumed to follow a Gaussian process prior, that is 
\begin{gather}
	f(\cdot)\sim GP(\mu(\cdot),k(\cdot,\cdot;\cuu{\theta}))
\end{gather}
where $\mu(\cdot)=E[f(\cdot)]$ is the mean function of Gaussian process, and $k(\cdot,\cdot;\cuu{\theta})$ is the covariance function with hyper-parameter $\cuu{\theta}$.

Given training data $\D = \{\cuu{x}_i, y_i,i=1,\ldots,n\}$, we have a multivariate Gaussian distribution that is
\begin{gather}
	\big(f(\cuu{x}_1),...,f(\cuu{x}_n)\big)\sim\N(\cuu{\mu}(\cuu{X}),K(\cuu{X},\cuu{X}))
\end{gather}
where $\cuu{X}$ refers to $(\cuu{x}_1,...,\cuu{x}_n)$, $\cuu{\mu}(\cuu{X}) =  (\mu(\cuu{x}_1),...,\mu(\cuu{x_n}))$ refers to a mean vector, and $K(\cuu{X},\cuu{X})$ refers to a $n\times n$ covariance matrix in which the entry of the $(i,j)$-th element is $k(\cuu{x}_i,\cuu{x}_j;\cuu{\theta})$.

To obtain the predictive distribution at a new input $\cuu{x}^*$, it is convenient to take advantage of the properties of Gaussian distribution:
\begin{gather}
	\begin{pmatrix}
		\cuu{y}\\
		\cuu{f}(\cuu{x}^*)
	\end{pmatrix}
	\sim\N
	\begin{pmatrix}
		\begin{pmatrix}
			\cuu{\mu}(\cuu{X}) \\
			\cuu{\mu}(\cuu{x}^*)
		\end{pmatrix}
		,
		\begin{pmatrix}
			K(\cuu{X},\cuu{X})+\sigma^2\cuu{I} &K(\cuu{x}^*,\cuu{X})^T\\
			K(\cuu{x}^*,X) &K(\cuu{x}^*,\cuu{x}^*)
		\end{pmatrix}
	\end{pmatrix}
\end{gather}
where $K(\cuu{x}^*,\cuu{X})$ is a $1\times n$ covariance matrix, $K(\cuu{x}^*,\cuu{x}^*)$ is scalar and their entries are similar to $K(\cuu{X},\cuu{X})$. Fortunately, the predictive distribution has an analytical form which is
\begin{gather}
	\begin{split}
		p(\cuu{f}(\cuu{x}^*)|\mathcal{D},\cuu{x}^*)\sim\N(&K(\cuu{x}^*,\cuu{X})^T(K(\cuu{X},\cuu{X})+\sigma^2\cuu{I})^{-1}(\cuu{y}-\cuu{\mu}(\cuu{X}))+\cuu{\mu}(\cuu{x}^*),\\
		&K(\cuu{x}^*,\cuu{x}^*)-K(\cuu{x}^*,\cuu{X})^T(K(\cuu{X},\cuu{X})+\sigma^2\cuu{I})^{-1}K(\cuu{x}^*,\cuu{X}))
		\label{eq:GP_predictive_distribution_3}
	\end{split}
\end{gather}
For more details of Gaussian process regression, see \cite{williams2006gaussian}.

The kernel function in the covariance function can be chosen from a parametric family, such as the radial basis function (RBF) kernel:
\begin{gather}
 	K(\cuu{x},\cuu{x}';\cuu{\theta})=\lambda_1^2\exp\bigg(-\frac{\|\cuu{x}-\cuu{x}'\|^2}{2\lambda_2}\bigg)
\end{gather}
with the hyper-parameter $\cuu{\theta}=\{\lambda_1,\lambda_2\}$ chosen by maximizing the marginal likelihood. Specifically, the hyper-parameters can be estimated by conjugate gradient descent algorithm \cite{press2007numerical} or Markov Chain Monte Carlo \cite{shi2011gaussian}.

\subsubsection{Wrapped Gaussian Distribution}

In this subsection, we first review the definitions of the wrapped Gaussian distribution and wrapped Gaussian process on a Riemannian manifold $\mathcal{M}$, and then review wrapped Gaussian process regression (WGPR)  which is introduced by \cite{mallasto2018wrapped}.

Suppose $\M$ denotes a $d$-dimensional \rma\ and $X$ is a random variable on $\M$. For $\mu \in\M$ and a symmetric positive definite matrix $K\in \mathbb{R}^{d\times d}$, we can define a wrapped Gaussian distribution as follows
\begin{gather}
    X =\Exp(\mu,\cuu{v}),\ \cuu{v} \sim\N(\cuu{0},K)
\end{gather}
where $\mu$ is called \xie{basepoint} and $\cuu{v}$ is a vector belongs to the tangent space at  $\mu$ which follows a multi-variate Gaussian distribution with zero mean and covariance matrix $K$ (see Section 3.1 in \cite{mallasto2018wrapped}). 
The wrapped Gaussian distribution is formally denoted as
\begin{gather}
    X\sim\N_\M(\mu,K).
\end{gather}

\cite{mallasto2018wrapped} describe a manifold-valued Gaussian process constructed by the wrapped Gaussian distribution, and the following material is based on their work. A collection $(f(X_1),...,f(X_n))$ of random points on a manifold $\M$ index over a set $\Omega$ is a wrapped Gaussian process, if every subcollection follows a jointly wrapped Gaussian distribution on $\M$.
It is denoted as 
\begin{gather}
    f(\cdot)\sim \mathcal{GP}_\M(\mu(\cdot),k(\cdot,\cdot;\cuu{\theta}))
\end{gather}
where  $\mu(\cdot)=E[f(\cdot)]$ is the mean function on a Riemannian manifold, and $k(\cdot,\cdot;\cuu{\theta})$ is a covariance function with hyper-parameter $\cuu{\theta}$ on a tangent space at  $\cuu{\mu}(\cdot)$.

Consider the wrapped Gaussian process regression model which relates the manifold-valued response $p\in \M$ to the vector-valued predictor $\cuu{x}\in \mathbb{R}^Q$ though a link function $f:\mathbb{R}^Q\rightarrow \M$ which we assume to be a wrapped Gaussian process.  
Analogously to Gaussian process regression, the authors also derive inferences for wrapped Gaussian process regression in Bayesian framework. Specifically,
given the training data $\mathcal{D}_\M=\{(\cuu{x}_i,p_i)|\cuu{x}_i\in \mathbb{R}^Q,p_i\in\M, i=1,\cdots,n\}$, the prior for $f(\cdot)$ is assumed to be a wrapped Gaussian process, which is $f(\cdot)\sim \mathcal{GP}_\M(\mu(\cdot),K(\cdot,\cdot;\cuu{\theta}))$. Then the joint distribution between the training outputs $\cuu{p}=(p_1,\cdots,p_n)$ and the test outputs $ {p}^*$ at $\cuu{x}^*$ is given by
\begin{gather}
\left(\begin{array}{c}
	 {p}^{*} \\
	\cuu{p}
\end{array}\right) \sim \mathcal{N}_{\M}\left(\left(\begin{array}{c}
	\mu^{*} \\
	\cuu{\mu}
\end{array}\right),\left(\begin{array}{cc}
	K^{* *} & K^{*} \\
	K^{*\top} & K
\end{array}\right)\right)
\end{gather}
where $\mu^*=\mu(\cuu{x}^*)$, $\cuu{\mu} = (\mu(\cuu{x}_1),...,\mu(\cuu{x_n}))$, $K=K(\cuu{x},\cuu{x})$,
$K^{*}=K(\cuu{x},\cuu{x}^*)$,     $K^{* *}=K\left(\boldsymbol{x}^{*}, \boldsymbol{x}^{*}\right)$ with $\cuu{x}=(\cuu{x}_1,\cdots,\cuu{x}_n)$.  

Therefore, by Theorem 1 in \cite{mallasto2018wrapped}, the predictive distribution for new data given the training data can be obtained by
\begin{gather}
{p}^{*}  \mid 	\cuu{p}  \sim \operatorname{Exp} \left(\mu^*,v^*\right), 
\end{gather}
where
\begin{gather}
	v^*\sim \mathcal{N}_\mathcal{M}\left(m^{*}, \Sigma^{*}\right),\ m^{*}=K^{*} K^{-1} \Log(\cuu{\mu},\cuu{p}),\ \Sigma^{*}=K^{* *}-K^{*} K^{-1} K^{*\top}.
\end{gather}

Similar to the comparison of operators between Riemannian manifolds and Euclidean spaces, Table \ref{tab:GPR_and_WGPR} shows the analogs of Gaussian processes between these two spaces.

\begin{table}[h]

\centering
    \caption{Analogs between Gaussian processes in Euclidean spaces and Riemannian manifolds.}
    \begin{tabularx}{\textwidth}{Y Y Y}
        \hline
        &Euclidean Spaces &Riemannian Manifolds\\
        \hline
        Definition  &$f(\cdot)\sim GP(\mu(\cdot),k(\cdot,\cdot;\cuu{\theta}))$ &$ f(\cdot)\sim \mathcal{GP}_\M(\mu(\cdot),k(\cdot,\cdot;\cuu{\theta}))$  \\
            Mean &$\mu(\cdot) \in \R$ & $\mu(\cdot) \in \M$ \\
            Covariance &$k(\cdot,\cdot;\cuu{\theta}) \in \R^{Q\times Q}$ &$k(\cdot,\cdot;\cuu{\theta}) \in T_{\mu(\cdot)}\M^{Q\times Q}$
        \\
        \hline
    \end{tabularx}
\label{tab:GPR_and_WGPR}
\end{table}

\section{Main Model and Inference}
\label{sec:model}

Gaussian process regression is a powerful non-parametric tool in statistics and machine learning. In Section \ref{sec:model}, we adapt GPR to functional data on manifolds  by proposing the WGPFR model for batch data on \rma. Inference of the WGPFR model is presented in Section \ref{inferencewgpfr}, where the statistical properties of the model parameters are studied and an efficient algorithm for inference is proposed. 

\subsection{Wrapped Gaussian Process Functional Regression Model}

The data consist of discrete observations of $M$ random functional curves on $\M$, so that the $m$-th curve is denoted as $y_m(t)$, $t \in \T$, $m=1, \ldots, M$. We assume that all curves are observed at a set of times $t_i$, $i=1,\ldots,N$. Throughout, $i$ will index time and $m$ will index curves. 
The set of observations is therefore $y_{mi}=y_m(t_{i})$, $i=1,\ldots,N$, $m=1,\ldots,M$. Associated with the $m$-th curve, we assume there is a real vector-valued functional covariate $\vesub{x}m(t)\in \mathbb{R}^Q$. We assume, as for the functional responses $y_m(t)$, that  $\vesub{x}m(t)$ is also observed at the times $t_{i}$, so that we are given points $\vesub{x}m=\vesub{x}m(t_{i})$. Finally, we assume there is a vector-valued covariate $\vesub{u}{m} \in \mathbb{R}^P$ observed for each curve $y_m$. Thus, the data for the $m$-th curve comprises $y_{mi}, \vesub{x}m, t_{i}$, $i=1,\ldots, N$ together with the vector $\vesub{u}{m}$. These data, for fixed $m$, are called a \emph{batch}. 

\begin{figure}[h]
    \centering
    \includegraphics[width=0.8\textwidth,trim={0cm 0cm 0cm 0cm},clip]{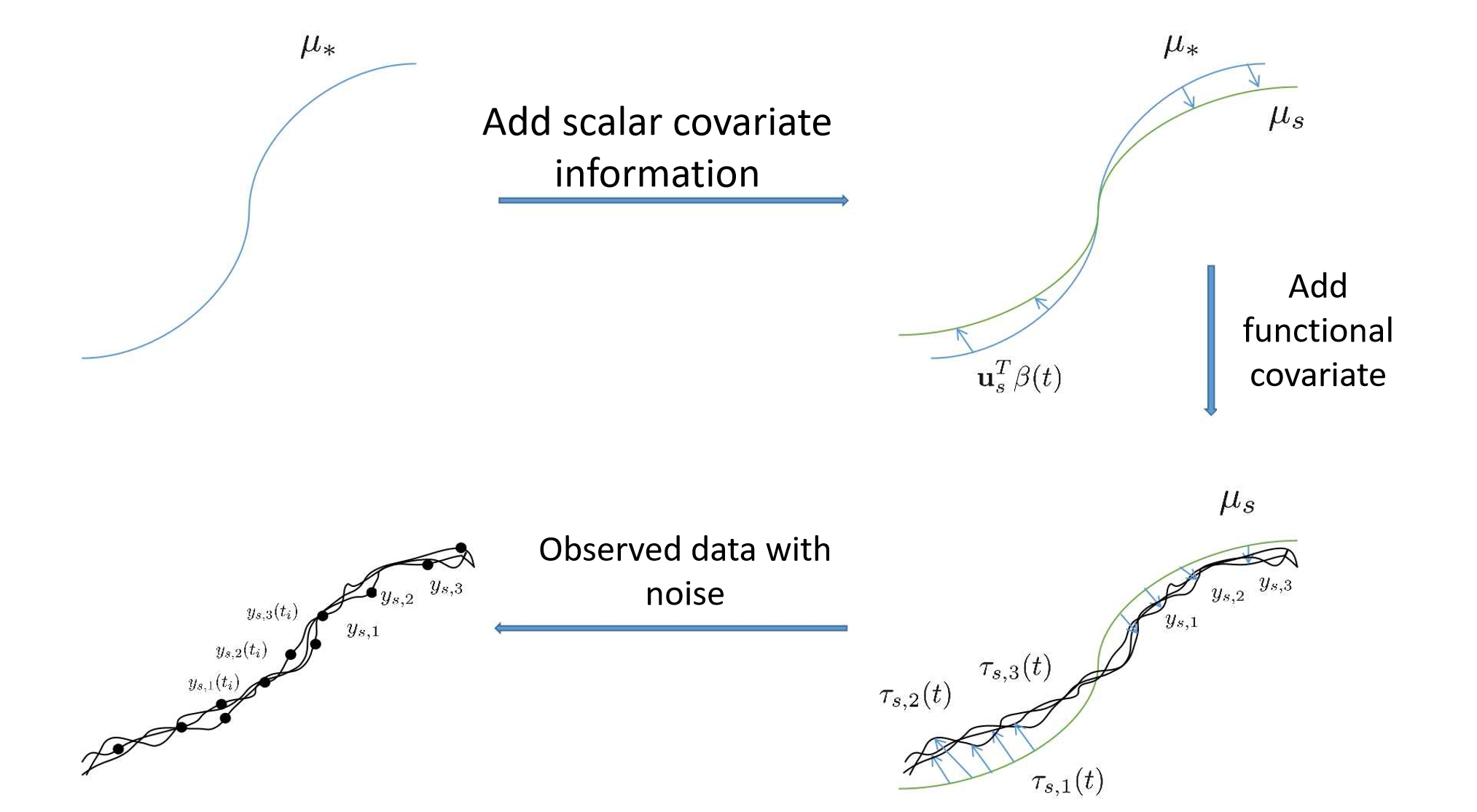}
    \caption{Schematic diagram of the model.}
\label{fig:schematic}
\end{figure}

Next we define our hierarchical model for the data. It has the following overall structure, and details of each part are filled out in what follows. At the top level of the hierarchy, there is a mean curve $\mu_*(t)\in\M$. There are two stages to obtain each curve $y_m$ from $\mu_*$. First there is a perturbation of $\mu_*$ to obtain an unobserved curve $\mu_m$. We assume this perturbation depends on the vector covariate $\vesub{u}{m}$ and a functional model parameter $\cuu{\beta}(t)\in \mathbb{R}^P$, but not on the functional covariates (see Equation~\eqref{eq:mean_structure}). Secondly, we assume $y_m$ is obtained from $\mu_m$ via a vector field $\tau_m$ along $\mu_m$ which depends on the functional covariate $\vesub{x}m(t)$ (see Equation~\eqref{eq:model}). Figure~\ref{fig:schematic} shows a schematic of the overall model. 

\subsubsection{Definition of Mean Structure}

It is difficult to infer non-linear relationships non-parametrically between a functional response variable and multi-dimensional functional covariates even when both of them are observed in a space for real-valued functions \cite{shi2007gaussian}. Therefore, we assume that the mean structure $\mu_m(\cdot) $ depends on the scalar covariates $\cuu{u}_m$ only. Specifically we assume 
\begin{gather}
	\mu_m(t) = \Exp(\mu_*(t), \cuu{u}_m^T\cuu{\beta}(t))
	\label{eq:mean_structure}
\end{gather}
where $\mu_*(t)\in\M$ and $\cuu{\beta}(t)\in T_{\mu_*(t)}\M$ play  a similar role to the intercept term and slope term of the function-on-scalar linear regression model in Euclidean space. 
These are model parameters which we estimate. Furthermore, equation \eqref{eq:mean_structure} can be seen as a functional version of geodesic regression proposed in \cite{fletcher2013geodesic}, in which a geodesic regression model for manifold-valued response and scalar-valued predictor is proposed in the form $y=\Exp(p,xv)$ with $(p,v)\in \mathcal{TM}$ where $x\in \mathbb{R}$ is the predictor variable.

\subsubsection{Definition of Covariance Structure}

Conditional on $\mu_m$, the $m$-th curve $y_m$ is modelled as
\begin{gather}
	y_m(t)=\Exp(\mu_m(t),\tau_m(t)), \text{for}\ m=1,...,M
	\label{eq:model}
\end{gather}
where $\tau_m(t) \in T_{\mu_m(t)}\M$ is a vector field along $\mu_m$. When the data are observed with random noise, \cite{fletcher2013geodesic} defines the geodesic model as $y=\Exp(\Exp(p,xv),\epsilon)$. Similarly, the proposed WGPFR model with error term can be defined as  
\begin{gather}
	y_m(t)=\Exp(\mu_m(t),\tau_m(t)+\epsilon_m(t)),    \text{for}\ m=1,\cdots,M;
	\label{eq:model add error}
\end{gather}
where $\epsilon_m(t)$ are independent measurement errors following a multivariate normal distribution with mean zero and covariance matrix $\sigma^2\mathbf{I}$. For functional data, in addition to the measurement error term, the most important part we are interested is the underlying covariance structure contained in term $\tau_m(t)$, which is Equation \eqref{eq:model}. In what follows, we explain how $\tau_m(t)$ is modelled as depending on the functional covariate $\vesub{x}m(t)$.  

We assume that the vector field $\tau_m(t)\in T_{\mu_m(t)}\M$ can be represented in local coordinates as a $d$-dimensional vector function $\tau_m(t)=(\tau_{m,1}(t),...,\tau_{m,d}(t))\in\R^d$. 
We further assume this vector-valued function follows a Gaussian process with mean zero and an underlying covariance structure $k_m(\cdot,\cdot;\cuu{\theta})$, which is
\begin{gather}
    \tau_m(t) \sim GP(\mathbf{0}, k_m(\cdot,\cdot;\cuu{\theta})). 
\end{gather}
Specifically, given two observation times $t_{i}$ and $t_{mj}$ from the $m$-th batch of data, we assume that the covariance of the two random vectors $\tau_m(t_{i})$ and $\tau_m(t_{mj})$ a function of the observed covariate values $\vesub{x}m(t_{i})$ and $\vesub{x}m(t_{mj})$.

Under the framework of the wrapped Gaussian process in \cite{mallasto2018wrapped}, we can represent the above Gaussian process as the wrapped Gaussian process for manifold functional data. That is, 
\begin{gather}
	y_m(t)\sim \mathcal{GP}_\M(\mu_m(t),k_m(\cdot,\cdot;\cuu{\theta}))
\end{gather}
where $\mu_m(t)\in \M$ and $k_m(\cdot,\cdot;\cuu{\theta})\in T_{\mu_m(t)}\M$ are defined as above. 

To summarize, the WGPFR model studies the non-linear relationship between the manifold-valued functional response and mixed (functional and vector) Euclidean-valued predictors in two parts, as shown in Figure~\ref{fig:schematic}. The first part is the mean structure which is defined through a functional regression to model the manifold-valued functional mean with the vector-valued predictor. The second part is called the covariance structure, and for which the inference is based on a Bayesian method with the prior assumed to be Gaussian process.

Compared to the geodesic regression in \cite{fletcher2013geodesic}, the WGPFR additionally incorporates the information of dependence between data modelled by covariance structure which is the main focus of interest for functional data analysis. Compared to the WGPR model in  \cite{mallasto2018wrapped}, this paper further models the mean part which relates the functional mean on the manifold to the scalar predictor on Euclidean space, while the mean part in \cite{mallasto2018wrapped} is not related to scalar predictors. Note that the scalar predictor plays an important role in the prediction of the model, especially for the batch data we consider in this paper. The proposed WGPFR model is more complicated but much flexible than existing models and induces a more complex inference procedure which will be illustrated in detail in the next subsection.

\subsection{Inference for WGPFR}
\label{inferencewgpfr}

\subsubsection{Estimation of Mean Structure}
\label{sec:estimated_mean_structure}

Since the mean structure model \eqref{eq:mean_structure} can be seen as a functional version of the geodesic regression model \cite{fletcher2013geodesic}, an intuitive inference method consists of optimizing $\mu_*(t) $ and $\cuu{\beta}(t)$ simultaneously for each fixed $t$. However, this would ignore information about the correlation of the function $\mu_*$ and $\cuu{\beta}(t)$ at different times. On the other hand, if the number of observed times $t_{i}$ is large, and we account fully for correlations between all different times, then the computational cost will be very high, unless we use some efficient methods, such as \cite{datta2016hierarchical}. Due to the dependent structure, we cannot conduct optimization at each fixed $t$ separately, otherwise the correlation information across different t will be lost. It can also be seen that inference of the functional mean structure model is indeed challenging.  

In this paper we borrow an idea of \cite{dai2018principal} to estimate $\mu_*(t) $ and $\cuu{\beta}(t)$.  Note that $\mu_*(t) $ plays the role of an intercept in the linear regression in Euclidean space, which is usually defined as the mean of the response. Similarly, for functional data $y_m(t), m=1,...,M$ on a \rma, we can define the intrinsic population Fr\'{e}chet mean function under the assumption of existence and uniqueness:
\begin{equation}
	\mu_0(t)=\underset{p(t)\in\M}{\text{arg min}}\, E[d_\M(y(t),p(t))^2]
	\label{eq:intrinsic_frechet_mean}
\end{equation}
where $d_\M(\cdot,\cdot)$ denotes geodesic distance. When $t$ is fixed, $\mu_0(t)$ refers to a point on a \rma; when $t$ is variable, $\mu_0(t)$ refers to a curve on a \rma. Additionally, since each $y_{m}(t) \in \M$ is continuous, $\mu_0(t)$ is also continuous (see equation (1) in \cite{dai2018principal}).

The intrinsic Fr\'{e}chet mean function $\mu_0(t)$ can be a good choice for $\mu_*(t)$ since it is a natural mean of manifold-valued functional data and it is reasonable to assume $y_m(t), m=1,...,M$ are confined to stay within the radius of injectivity at  $\mu_0(t)$ for all $t\in \mathcal{T}$. Then the inverse exponential map can be  well-defined. We estimate $\mu_*(t)$ at each time $t$ via the sample Fr\'echet mean of the points $y_m(t)$. This requires that the times $t_{i}$ are the same across all curves $m=1,\ldots,M$. Standard gradient descent methods \cite{pennec1999probabilities} can be used to obtain the Fr\'echet sample mean at each time $t=t_i$ to obtain estimates $\hat{\mu}_0(t_i)$ of $\mu_{*}(t_i)$, $i=1,\ldots,N$.  
Denote the logarithm process data as
$$
V_m(t)=\Log(\mu_0(t),y_m(t)),
$$ 
where $V_m(t)$ is a $d$-dimensional tangent vector of $y_m(t)$ at $\mu_0(t)$ for a fixed $t$. 
Since we assume $\M$ is a $d$ dimensional closed Riemannian submanifold of a Euclidean space,  it is reasonable to identify the tangent space $T_{\mu_0(t)}\M$  as a subspace of the Euclidean space $\mathbb{R}^{d_0}$ with $d_0\geq d$. Then $V_m(t)$ can be seen as $\mathbb{R}^{d_0}$ valued square integrable functions, which can be represented by a set of basis functions. Similarly, the slope function $\cuu{\beta}(t)$ can also be represented by this set of basis functions. In this way, we can not only easily estimate the intercept term $\mu_0(t)$ but also the slope term $\cuu{\beta}(t)$ can be obtained by least square methods on $\mathbb{R}^{d_0}$. This reduces the computation complexity compared with the optimization method in \cite{fletcher2013geodesic}. Next we illustrate the estimation procedure in detail.
 
For the estimation of the intrinsic Fr\'{e}chet mean $\mu_0(t)$, the sample Fr\'{e}chet mean is calculated by minimizing the following loss function for each $t\in \mathcal{T}$:
\begin{gather}
    \label{samplemean}
    \hat{\mu}_0(t) = \underset{p(t)\in\M}{\text{arg min}}\ \frac{1}{M}\sum_{m=1}^{M}d_\M(y_{m}(t), p(t))^2.
 \end{gather}
It is natural to use the gradient descent algorithm introduced by \cite{pennec1999probabilities} to estimate the sample Fr\'{e}chet mean for each fixed $t$. Using \cite{do1992mathematics}, the gradient is given by 
\begin{gather}
	\nabla d_\M(p,y_m(t))^2 = -2\Log(y_m(t),p).
\end{gather}
An implementation of the gradient descent algorithm to solve equation (\ref{samplemean}) is given in Algorithm \ref{alg:GDA_sample_frechet_mean}. In practice, we choose the step size $l=\frac{1}{N}$ where $N$ is the number of samples and $\epsilon=10^{-8}$ which is relatively small.

\begin{algorithm}[H]
    \label{alg:GDA_sample_frechet_mean}
	\caption{Gradient Descent Algorithm for Sample Fr\'{e}chet Mean, fixed $t$.}
	\SetAlgoLined
	1. Initialise with a manifold-valued point $p$ and set a step size $l$ and a small positive number $\epsilon$ \;
	2. $\nu = \frac{2l}{M}\sum_{m=1}^M\Log(y_m(t),p)$ \;
	3. While $\Vert\nu\Vert > \epsilon$, do $p=\Log(p,\nu)$; else, end.
\end{algorithm}
 
For the estimation of  $\cuu{\beta}(t)=\big(\cuu{\beta}_1(t),\cdots,\cuu{\beta}_p(t)\big)$, the first step is to find a reasonable basis to represent $\cuu{\beta}(t)$ in the tangent space. Under the framework of \cite{dai2018principal}, suppose we have an arbitrary system of $K$ orthonormal basis functions
\begin{gather}
     \Phi_K=\{\phi_k\in \mathbb{H}~|~\phi_k(t)\in T_{\mu_0(t)}\M,\langle \phi_k,\phi_l \rangle=\delta_{kl},k,l=1,\cdots,K\},
\end{gather}
where $\mathbb{H}=\{\nu:\mathcal{T}\rightarrow\mathbb{R}^d,\int_{\mathcal{T}}\nu(t)^T\nu(t)dt<\infty\}$ is a $L^2$ Hilbert space of $\mathbb{R}^d$ with inner product $\langle \nu,u \rangle=\int_{\mathcal{T}}\nu(t)^Tu(t)dt\}$ and norm $\Vert v\Vert=\langle v,v \rangle^{1/2}$, $\delta_{kl}=1$ if $k=l$ and 0 otherwise. Note that the values of each $\phi_k(t)$ at each time $t\in \mathcal{T}$ is restricted to the tangent space $ T_{\mu_0(t)}\M$, which are identified with $\mathbb{R}^{d_0}$. Define the $K$-dimensional linear subspace of $\mathbb{H}$:
\begin{gather}
 \mathcal{M}_{K}\left(\Phi_{K}\right):=\left\{ x(t)=\sum_{k=1}^{K} a_{k} \phi_{k}(t) \text { for } t \in \mathcal{T} \mid a_{k} \in \mathbb{R}\right\}.
\end{gather}
Then the slope functions  $\cuu{\beta}_j(t)\in T_{\mu_0(t)}\M$, $j=1,\cdots,p$  can be approximated by a linear span on $ \mathcal{M}_{K}\left(\Phi_{K}\right)$ with expansion coefficients  $b_{jk}$: 
\begin{gather}
    \cuu{\beta}_j(t)\approx \sum_{k=1}^{K} b_{jk} \phi_{k}(t).
\end{gather}
Then the estimation of $\cuu{\beta}_j(t)$ can be transformed to the estimation of $ b_{jk}$, $k=1,2,\cdots,K$, $j=1,2,\cdots,p$.

Defining $V_m(t)=\Log(\mu_0(t),y_m(t))$, $\hat{V}_m(t)=\Log(\hat{\mu}_0(t),y_m(t))$ and
$W_m(t) =\Log(\mu_0(t),\mu_m(t))$, then the multiple linear function-on-scalar regression model becomes 
\begin{gather}
    V_m(t)=W_m(t)+e_m(t)\approx \sum_{j=1}^p\sum_{k=1}^{K}u_{mj}b_{jk} \phi_{k}(t)+e_m(t)
\end{gather}
where $e_m(t)=V_m(t)-W_m(t)$ which are assumed to be independent of the covariate $\cuu{u}_m$ for $m=1,\cdots,M$.

Since $V_m(t)$ can not be obtained directly, we use $\hat{V}_m(t)$ as the response of the multiple linear regression model, so that the parameters of the regression model are estimated by minimizing the following loss function with observed data:
\begin{gather}
    L(b_{jk})=\sum_{m=1}^M\sum_{i=1}^n\|\hat{V}_m(t_i)-\sum_{j=1}^p\sum_{k=1}^{K}u_{mj}b_{jk} \phi_{k}(t_i)\|^2.
\label{eq:estimated_b}
\end{gather}
The $b_{jk}$ coefficients are calculated by standard least squares methods.

To be more specific, let
$\hat{\cu{V}}=(\hat{V}_{im})_{i=1,\cdots,nd_0;m=1,\cdots,M}$ denote a $nd_0\times M$ matrix such that the $m$-th column denotes the $m$-th batch, and the $i$-th row denotes the vectored element $\bigg(\hat{V}_{m1}(t_1),\cdots,\hat{V}_{m1}(t_n),\cdots,\hat{V}_{md_0}(t_1),\cdots,\hat{V}_{md_0}(t_n)\bigg)$. Similarly, let 
$\cuu{\Phi}=(\phi_{ik})_{i=1,\cdots,nd_0;k=1,\cdots,K}$ denote a $nd_0\times K$ matrix; let $U$ denote a $M\times p$ matrix for which each row $\cu{u}_m$ indicates a data point; and let $B$ denote the $p\times K$ coefficient matrix with element $b_{jk}$ for $j=1,\cdots,p$, $k=1,\cdots,K$.  Then the above loss function can be rewritten as
\begin{gather}
    L(B)=\|vec(\hat{\cu{V}})-(\cuu{\Phi}\otimes U)vec(B)\|^2.
\end{gather} 
The least squares estimate of $B$ is obtained by
\begin{gather}
    vec(\hat{B})=\bigg((\cuu{\Phi}\otimes U)^\top(\cuu{\Phi}\otimes U)\bigg)^{-1}(\cuu{\Phi}\otimes U)^\top vec(\hat{\cu{V}})
\end{gather}

Consistency of the estimator of the sample Fr\'{e}chet mean was proved in \cite{dai2018principal}. The following theorem establishes consistency of the estimator of the regression coefficients; the proof is left to the appendix.

\begin{theorem}
    \label{theorem_1}
    Under the conditions (C1)-(C5) in the Appendix, the coefficient  $vec(\hat{B})$ is a consistent estimator with probability tending to $1$ as $n \rightarrow \infty$ in the sense that
    \begin{gather}
    	\|vec(\hat{B})-vec({B})\|=o_p(1).
    \end{gather}
\end{theorem} 

It follows that the functional slope coefficients can be estimated by  $\hat{\cuu{\beta}}(t)=\big(\hat{\cuu{\beta}}_1(t),\cdots,\hat{\cuu{\beta}}_p(t)\big)$ where
\begin{gather}
    \hat{\cuu{\beta}}_j(t)= \sum_{k=1}^{K} \hat{b}_{jk} \phi_{k}(t)
\end{gather}
for $j=1,\cdots,p$.
The estimated mean structure for the $m$-th curve is given by
\begin{gather}
	\hat{\mu}_{m}(t)=\Exp(\hat{\mu}_0(t),\cu{u}_m\hat{\cuu{\beta}}(t)).
	\label{eq:estimated_mean}
\end{gather}
The function $\hat{\mu}_{m}(t)$ is expected to approximate $\mu_m(t)$. However we omit the rigorous proof of the consistency of $\hat{\mu}_{m}(t)$ on the manifold due to some technique difficulties.

\subsubsection{Estimation of Covariance Structure}
\label{sec:estimate_covariance_structure}

From the section above, we obtain an estimate of the mean structure $\mu_m(t)$ which is a continuous function on $\M$. In this section, we will focus on inference of the covariance structure $\tau_m(t)\in T_{\mu_m(t)}\M$. We assume that the tangent spaces $T_{\mu_m(t)}\M$ for $t\in\mathcal{T}$ can be identified with $\R^{d_0}$ via some smooth local basis of orthonormal vector fields along $\mu_m(t)$. We have mentioned above that the covariance structure can be related to another functional covariate $x_m(t)$, which is a $\mathbb{R}^q$-valued function. If we denote $\tau_m(t)=\tau_{m}(x_m(t))=(\tau_{m1}(t),\cdots,\tau_{md_0}(t))$, then the correlation of different components in $\tau_m(t)$ could be estimated via a cross-covariance function model, such as the convolved Gaussian process \cite{boyle2004dependent}. However, if there are $n$ observations $\tau_{md}(t_i)$, $i=1,\cdots,n$, then the size of a covariance matrix in a Gaussian process of $\tau_{md}(t)$, $d=1,\cdots,d_0$ is $n\times n$ while the size of a cross-covariance matrix in a convolved Gaussian process of $\tau_{m}(t)$ is $n^{d_0}\times n^{d_0}$, which is computationally expensive. As a result, we will consider different dimensions independently. Specifically, we assume
\begin{gather}
	\tau_{md}(t)\sim GP(0,K_{md}(x_m(t),x_m(t');\cuu{\theta}_{md})),
\end{gather}
where $K_{md}(\cdot,\cdot;\cuu{\theta}_{md})$ denotes a covariance kernel depending on hyper-parameter $\cuu{\theta}_{md}$.

Given the value of $\tau_{md}(t_{i})$, $i=1,\cdots,n$, denote $\cuu{\tau}_{md}=(\tau_{md}(t_{m1}),...,\tau_{md}(t_{mn}))$, then for any new input $t_m^*$, the conditional distribution of $\tau_{md}(t_m^*)$ is
\begin{gather}
    \tau_{md}(t_m^*)|\D_{md} \sim \N(\mu_{md}^*, \Sigma_{md}^*),\ \mu_{md}^* = \cuu{k}_{md}^{*T}K_{md}^{-1}\cuu{\tau}_{md},\ \Sigma_{md}^* = k_{md}^{**} - \cuu{k}_{md}^{*T}K_{md}^{-1}\cuu{k}_{md}^*
\end{gather}
where $\cuu{k}_{md}^*=(K_{md}(t_{m1},t_m^*),...,K_{md}(t_{mn},t_m^*))$ and $k_{md}^{**} = K_{md}(t_m^*,t_m^*)$.

In Gaussian process regression, the hyper-parameters for the $d$-th dimension, $\cuu{\theta}_{md}$, can be estimated by maximizing the sum of marginal likelihood functions for each batch, i.e.~maximizing
\begin{gather}
    (2\pi)^{-\frac{n}{2}}\det(K_{md})^{-\half}e^{-\half\cuu{\tau}_{md}^TK_{md}^{-1}\cuu{\tau}_{md}}.
\label{eq:likelihood_WGP}
\end{gather}

Under some regularity conditions, the estimator $\hat{\theta}_{md}$ converges to $\theta_{md}$ \cite{choi2011gaussian}. With the estimated hyper-parameters, the estimated covariance structure of the $d$-th coordinate for the $m$-th batch is given by
\begin{gather}
    \hat{\tau}_{md}(t)|{\cuu{\tau}}_{md} \sim GP(\cuu{k}_{md}^T(t)K_{md}^{-1} {\cuu{\tau}}_{md}, K_{md}(t,t)-\cuu{k}_{md}(t)^{T}K_{md}^{-1}\cuu{k}_{md}(t)|\hat{\cuu{\theta}}_{md}))
	\label{eq:estimated_covariance_approx_model_rma}
\end{gather}
where $\cuu{k}_{md}(t)=(K_{md}(t,\cuu{x}_m(t_{m1})),\cdots,K_{md}(t,\cuu{x}_m(t_{mn})))$. Under some mild conditions, the above estimator of the covariance structure is information consistent, as the following theorem states.

\begin{theorem}
    \label{theorem_2}
    Suppose $(1)$ the underlying true vector field, $\tau_{m}(t)$ along $\mu_m(t)$ can be represented using a local basis by real-valued functions $\tau_{md}$, $d=1,\ldots,d_0$, and that each function has a Gaussian process prior with mean zero and bounded covariance function $K_{md}(\cdot,\cdot,\cuu{\theta}_{md})$ for any covariate $x_m(t)$, $(2)$ $K_{md}(\cdot,\cdot,\cuu{\theta}_{md})$ is continuous in $\cuu{\theta}_{md}$, further assume the estimator $\hat{\cuu{\theta}}_{md}$ converges to $\cuu{\theta}_{md}$ as $n\rightarrow \infty$ almost surely, and $(3)$ the mean structure $\mu_m(t)$ is known.
    
    Then $\hat{\tau}_{md}(\cdot)$ is information consistent to the true $\tau_{md}(\cdot)$, which means an estimator that guarantees convergence to the true parameter value as the sample size grows, if the reproducing kernel Hilbert space norm $\|\tau_{md}\|_k$ is bounded and the expected regret term  $E_{\boldsymbol{X}}\left(\log \left|\boldsymbol{I}+\sigma^{-2} \boldsymbol{C}_{n n}\right|\right)=o(n)$,  where $\boldsymbol{C}_{n n}=\left(k\left(\boldsymbol{x}_{i}, \boldsymbol{x}_{j} ; \boldsymbol{\theta}\right)\right)$ is the covariance matrix over $\boldsymbol{x}_{i}, i=1, \ldots, n$, and $\sigma^2$ is the variance of the measurement error $\epsilon_m(t)$.	
\end{theorem}

The proof is given in Appendix. Note that the above theorem holds under the condition that $\mu_m(t)$ is known, but in practice, we only have an estimate of $\mu_m(t)$. Furthermore, the term $\cuu{\tau}_{md}$ in the posterior distribution can not be observed directly, and it is approximated by $\tilde{\cuu{\tau}}_{md}=(\tilde{\tau}_{md}(t_{m1}),\cdots,\tilde{\tau}_{md}(t_{mn}))$ which is the realization of the $d$-th element of  $\tilde{\tau}_m(t)=(\tilde{\tau}_{m1}(t),\cdots,\tilde{\tau}_{md_0}(t))=\Log(\hat{\mu}_m(t),y_m(t))$.

\subsubsection{Update Mean Structure and Covariance Structure}
\label{sec:update}

After obtaining the estimated mean structure and covariance structure, we are able to make  predictions with given new inputs. In order to improve the performance of our model, we introduce an algorithm which can update the estimated mean structure and covariance structure iteratively.

The loss function of the $m$-th curve at time point $t_{i}$ with the estimated mean structure and estimated covariance structure is given as
\begin{gather}
E=\sum_{m=1}^M\sum_{i=1}^{n}d_{\M}\Bigg(\Exp\Big(\Exp\big(\hat{\mu}_0(t_{i}),\hat{\mu}_m(t)\big), \hat{\tau}_m(t_{i})\Big), y_m(t_{i})\Bigg)^2
	\label{eq:new_loss_given_cov}
\end{gather}
where $\hat{\mu}_m(t)=\sum_{j=1}^p\sum_{k=1}^Ku_{mj}\hat{b}_{jk}\phi_k(t_{i})$.

Given the above estimated covariance structure $\hat{\tau}_m(t_{i})$, the mean structure can be updated by a gradient descent algorithm where the gradient is
\begin{gather}
    \nabla_{\mu_m(t_{i})}E = -d_{\mu_m(t_{i})}\Exp\Big(\mu_m(t_{i}),\hat{\tau}_m(t_{i})\Big)^\dagger\Log\Big(y_m(t_{i}),\Exp\big(\mu_m(t_{i}),\hat{\tau}_m(t_{i})\big)\Big),
	\label{eq:estimate_mu_m}
\end{gather}
where $\dagger$ is an adjoint with respect to the Riemannian inner product, which plays a similar role to parallel transport. In practice, a variational method for gradient descent \cite{kim2014multivariate} can be used as a substitute for gradient in Equation \eqref{eq:estimate_mu_m}. Thereafter, we update the mean structure from $\hat{\mu}_m(t)$ to $\hat{\mu}_m^{(1)}(t)$ (here the superscript $^{(1)}$ means the $1$-st iteration). The updated coefficients $b_{jk}^{(1)}$ can be estimated by minimizing the loss function
\begin{gather}
    L(b_{jk})= \|\Log(\hat{\mu}_0(t_{i}), \hat{\mu}^{(1)}_m(t_{i}))-\sum_{j=1}^p	\sum_{k=1}^{K}u_{m,j}b_{jk}\phi_k(t)\|^2,
\end{gather}
which is a linear least squares problem.

Given the updated mean structure $\hat{\mu}_m^{(1)}(t)$, we can re-calculate the covariance structure $\hat{\tau}_m^{(1)}(t)$. The only difference from Section \ref{sec:estimate_covariance_structure} is that the estimated mean structure is replaced by the updated mean structure. This two-way updating procedure is then repeated iteratively, stopping when the difference of the updated mean or covariance structure smaller than some given threshold. The algorithm of the updating procedure is specified in Algorithm \ref{alg:approx_model_rma}.

\begin{algorithm}[h]
     \caption{Algorithm to estimate and update coefficients $B=(b_{jk})_{j=1,\cdots,p;k=1,\cdots,K}$ and hyper-parameters $\cuu{\Theta}=(\cuu{\theta}_{md})_{m=1,\cdots,M;d=1,\cdots,d_0}$.}
 	\SetAlgoLined
 	\KwIn{$y_{m}(t_{i}), t_{i},\cuu{x}_{m}(t_{i}), \cuu{u}_m, m=1,...,M, i=1,...,n,  iter=1$.}
 	\KwOut{$\hat{b}_{jk}$ and $\hat{\cuu{\Theta}}$.}
 	1. Compute sample Fr\'{e}chet mean function $\hat{\mu}_0(t)$ for all curves. \\
 	2. Estimate the coefficient $b_{jk}^{(0)}$ according to equation \eqref{eq:estimated_b}, to obtain the estimate of the mean structure $\hat{\mu}^{(0)}_m(t)$. \\
 	3. Estimate the hyper-parameter $\cuu{\Theta}^{(0)}$ by maximising the likelihood function of Equation~\eqref{eq:likelihood_WGP}, to obtain the estimate of the covariance structure $\hat{\tau}^{(0)}_m(t)$.\\
 	4. Update $\hat{\mu}_m^{(iter)}(t)$ according to Equation \eqref{eq:estimate_mu_m}.\\
 	5. Update $\cuu{\Theta}^{(iter)}$ with the updated  $\hat{\mu}^{(iter)}_m(t)$, and set $iter\ += 1$.\\
 	6. Repeat Steps 4 and 5 until some convergence conditions have been satisfied.
 	\label{alg:approx_model_rma}
\end{algorithm}

\section{Numerical Experiments}
\label{sec:numerical}

In this section we demonstrate the WGPFR model on two Riemannian manifolds: $S^2$ and Kendall's shape space. As previously, we suppose there are $M$ curves on a Riemannian manifold; in what follows we will simulate data points on these random curves in different ways to form different scenarios of interpolation and extrapolation. For the extrapolation problem, we use our model to predict the last $15$ data points (given all preceding data points) to test the long-term extrapolating performance, and the last $5$ data points to test the short-term extrapolation performance. Moreover, since the Algorithm \ref{alg:approx_model_rma} is calculated in Euclidean space, we use a optimizer from SciPy which is powerful and widely-used for minimization problems.

\subsection{Regression analysis on $S^2$}
\subsubsection{Simulation scenario on $S^2$}
\label{sec:general_model_data}

\begin{figure}[h]
    \centering
    \includegraphics[width=150mm,scale=0.5]{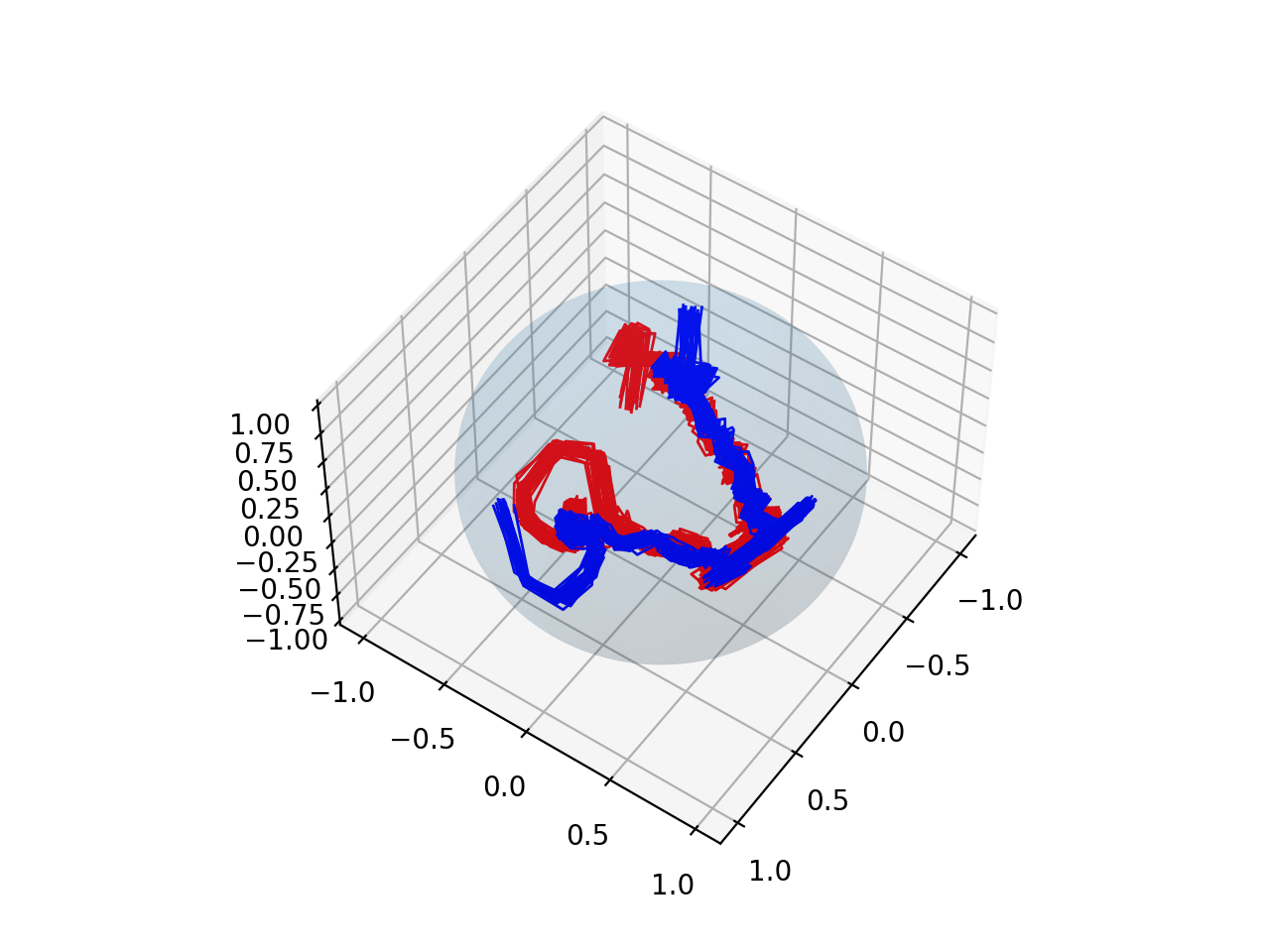}
    \caption{Simulated data on $S^2$ where red curves refer to one batch and blue curves refer to the other.}
\label{fig:general_model_data}
\end{figure}

Suppose observation times $t$ are equally spaced in the interval $[0, 1]$ with $N$ points, where $m\in\{1,...,m_1,m_1+1,...,M\}$ and $M=m_1+m_2$. To test the performance of our model with different number of observed curves and data points, we considered $30$ and $60$ batches and $20$, $40$ and $60$ data points on each batch respectively, which is $M \in \{30, 40\}$ and $n\in\{20,40,60 \}$. The simulated data are shown in Figure \ref{fig:general_model_data} and in the remainder of this subsection we explain how the data were generated.

As introduced previously, the WGPFR model consists of three parts: the intrinsic Fr\'{e}chet mean function, the mean structure and the covariance structure. In order to simulate data, we first define the Fr\'{e}chet mean function $\mu_0(t)$ on $S^2$ by
\begin{gather}
    \mu_0(t) = \Exp(p, (\sin{(\frac{t\pi}{2})^2}, \sin{(t\pi)^3,0)})
\end{gather}
where $p=(0,0,1)$ refers to a point on a $S^2$.

Suppose there are two batches, and that we generate two mean structures based on batch-specific covariates $\cuu{u}_m$ and Fourier basis functions $\cuu{\beta}(t)$, defined by
\begin{gather}
    \mu_m(t) = \Exp(\mu_0(t), \cuu{u}_s^T\cuu{\beta}(t)),\ m=1,...,m_1+m_2
\end{gather}
where $\cuu{u}_m=(1,0)$ for $m=1,...,m_1$ and $\cuu{u}_m=(1,1)$ for $m=m_1+1,...,m_1+m_2$ refer to the covariates for two scalar batches respectively and $\cuu{\beta}(t)=(\cuu{\beta}_1(t),\cuu{\beta}_2(t))$ refers to tangent vector-valued basis functions. Specifically, we define
\begin{gather}
\begin{split}
    \cuu{\beta}_1(t) &= R_t\big(\sum_{i=1}^{20} w_{1,i}(\phi_i(t), \phi_i(\frac{t+1}{2}), \phi_i(\frac{t+2}{2}))\big), \\
    \cuu{\beta}_2(t) &= R_t\big(\sum_{i=1}^{20} w_{2,i}(\phi_i(t), \phi_i(\frac{t+1}{2}), \phi_i(\frac{t+2}{2}))\big),\ i=1,...,20
\end{split}
\end{gather}
where
\begin{gather}
\begin{split}
    w_{1,i} &= \frac{i}{120},\\
    w_{2,i} &= -\frac{1}{2}\sqrt{\sin{(\frac{i}{60})}},\ i=1,...,20,\\
    \phi_i(t) &= \frac{1}{\sqrt{2}}, \ \text{for}\ i=0,\\
    \phi_i(t) &= \sin{(2\pi it)}, \ \text{for}\ i = 1,3,5,...,19,\\ \phi_i(t) &= \cos{(2\pi it)},\ \text{for}\ i = 2,4,6,...,20,\\
\end{split}
\end{gather}
and $R_t$ is a rotation matrix transferring the vector $\sum_{i=1}^{20} w_{1,i}(\phi_i(t), \phi_i(\frac{t+1}{2}), \phi_i(\frac{t+2}{2})))$ or 
\newline
$\sum_{i=1}^{20} w_{2,i}(\phi_i(t), \phi_i(\frac{t+1}{2}), \phi_i(\frac{t+2}{2})))$ to $T_{\mu_0(t)}\M$, this ensures that the vectors $\cuu{\beta}_1(t),\cuu{\beta}_2(t)$ are tangents at $\mu_0(t)$. For example, given a point $p$ and its tangent vector $\mathbf{v}$, the rotation matrix is defined by
\begin{gather*}
R_t(\mathbf{v})=\frac{\mathbf{v}-\langle v,p\rangle}{\Vert\mathbf{v}-\langle v,p\rangle\Vert}\Vert\mathbf{v}\Vert.
\end{gather*}
We use the following covariance function
\begin{gather}
    k(t_i, t_j) = v_0\exp(-\frac{1}{2w_0} (t_i-t_j)^2)+a_0+a_1t_it_j+\sigma^2\delta_{ij}
    \label{eq:kernel_function}
\end{gather}
which is a combination of a squared exponential kernel (often used for stationary Gaussian processes) and a linear kernel (often used for non-stationary Gaussian process), this can be used for most of Gaussian process regression \cite{shi2007gaussian}. Since $S^2$ can be embedded into $\R^3$, for each dimension of tangent vector of covariance structure, the hyper-parameters $(v_0,w_0,a_0,a_1,\sigma)$ are given as $\cuu{\theta}_1=(0.012, 3, 0.01, 0.01, 0.02)$, $\cuu{\theta}_2=(0.017, 3.1, 0.011, 0.012, 0.015)$ and $\cuu{\theta}_3=(0.015, 3.2, 0.012, 0.013, 0.01)$. Given these three Gaussian processes with zero mean and kernel \ref{eq:kernel_function}, we can generate a tangent vector for the covariance structure. However, in practice, the tangent vector may not in the target tangent space because of numerical rounding. For example, if the ideal tangent vector is $(1,1,0)$, our generated tangent vector might be $(0.99,1,0.01)$, which will result in the exponential map and inverse exponential map incalculable (defined in Equation \eqref{eq:exp_sphere} and Equation \eqref{eq:log_sphere} respectively). Therefore, the tangent vector must be projected into the correct tangent space, which is denoted as $v_3(t)|\cuu{\theta}_1,\cuu{\theta}_2,\cuu{\theta}_3$. The subscript 3 in $v_3$ refers to $3$-dimensional. As a consequence, the formula to generate manifold-valued data for numerical experiments is written as
\begin{gather}
    y_{s,m}(t)=\Exp(\mu_s(t),v_3(t)|\cuu{\theta}_1,\cuu{\theta}_2,\cuu{\theta}_3),\ s=1,2,\ m=1,...,M_s.
\label{eq:generate_general_data}
\end{gather}

\subsubsection{Model assessment on $S^2$}

In this section, we delete some generated data points on a randomly selected curve, say $y_r(t)$, in different ways to form training data sets and then calculate the predictions for these deleted data points (i.e. the test data). The performance of WGPFR model is assessed by comparing the root mean square error between predictions and their real data by Euclidean distance, since a sphere can be embedded in $\mathbb{R}^3$ easily.

Specifically, we selected $15$ data points on $y_r(t)$ uniformly at random as our test data set, and all remaining data points were used as the training data set, which is a typical interpolation problem. This scenario is denoted as Type \rom{1} prediction. As to Type \rom{2} prediction, the short-term forecasting, the last $5$ data points on $y_r(t)$ are considered as test data set which is a typical extrapolation problem. Analogously to the short-term forecasting, we also test the performance of WGPFR for long-term forecasting, the Type \rom{3} prediction. In Type \rom{3} prediction, we choose the last $15$ data points as test data. 

For comparison, in each scenario, the same training data set is used for several other models, such as functional linear regression on Riemannian manifolds (FLRM) and wrapped Gaussian process Fr\'{e}chet mean regression (WGFmR). Specifically, FLRM is the mean structure \eqref{eq:mean_structure} without covariance structure; WGPFmR consists of mean structure and covariance structure in which the mean structure is the sample Fr\'{e}chet mean point for all training data. In addition, the WGPFmR model does not have the updating part. We compare the performance of these three models not only to show a significant improvement of predictive accuracy from concurrent model (consider mean structure and covariance structure simultaneously), but also from the intervatively updating algorithm \ref{alg:approx_model_rma}. Moreover, we can use some models, such as Gaussian process functional regression without manifold structure to fit the data and make predictions. However, such models are meaningless since the inferences cannot be guaranteed on the right space which is manifold. Thus, it might be not suitable as a baseline and we only consider comparison with manifold structure.

We replicate each simulation study $100$ times. Thus, we test the performance of our model on thousands of test data points. The numerical results reported in Table \ref{tab:rmse_general_model_equally_spaced} are the average of root-mean-square-error (RMSE) in every single replication. Using the embedding $S^2 \subseteq \R^3$, it is reasonable to use the Euclidean norm between points as a distance function (chordal metric), which provides a method to calculate the RMSE. 

In Type \rom{1} prediction, Type \rom{2} prediction and Type \rom{3} prediction, WGPFR always provides the best prediction. The predictive accuracy of FLRM is better than that of WGPFmR in most scenarios, because FLRM only learns the mean structure of the training data which might be more useful for trend forecasting. In extrapolation, the long-term forecasting and short-term forecasting, since test data are distant from the training data, the output of GRP is usually inaccurate and then the mean structure mainly determines the accurate of prediction. Because we used the Fr\'{e}chet mean point as the mean function of WGFmR, the prediction of WGFmR is very poor. 

Comparing long-term and short-term forecasting, since the test data of the former is closer to the training data than that of the latter, the predictions of short-term forecasting are more accurate than that of long-term forecasting. In addition, from the table we can see that when the numbers of curves are fixed, the RMSE between prediction and real data decreases with the increasing number of points; when the number of points is fixed, the RMSE also decreases with the increasing number of curves. 

\begin{table}[h]
    \centering
    \begin{tabularx}{\textwidth}{Y Y Y Y}
    \hline
          &Type \rom{1} &Type \rom{2} &Type \rom{3} \\
    \hline
    30 curves, 20 points\\
    \hline
    WGPFR &\textbf{0.0349} &\textbf{0.0226} &\textbf{0.0341} \\
    FLRM  &0.2194 &0.2951 &0.3188 \\
    WGFmR &0.2755 &0.3336 &0.4779 \\
    \hline
    30 curves, 60 points\\
    \hline
    WGPFR &\textbf{0.0209} &\textbf{0.0108} &\textbf{0.0248} \\
    FLRM  &0.1520 &0.2282 &0.2311 \\
    WGFmR &0.1909 &0.1608 &0.2164 \\
    \hline
    60 curves, 20 points\\
    \hline
    WGPFR &\textbf{0.0391} &\textbf{0.0273} &\textbf{0.0428} \\
    FLRM  &0.2117 &0.2509 &0.2865 \\
    WGFmR &0.2827 &0.3165 &0.4717 \\
    \hline
    60 curves, 60 points\\
    \hline
    WGPFR &\textbf{0.0215} &\textbf{0.0115} &\textbf{0.0126} \\
    FLRM  &0.1601 &0.2022 &0.2385 \\
    WGFmR &0.2058 &0.2199 &0.3225 \\
    \hline
    \end{tabularx}
    \caption{Root-mean-square-error for several models with four types of predictions and equally spaced data.}
    \label{tab:rmse_general_model_equally_spaced}
\end{table}

Since Gaussian process regression is a Bayesian model, we also compare the log-predictive likelihood of the covariance structure which provides the randomness of this model. \cite{gelman2014understanding} introduce a calculation of log pointwise predictive density in practice. Thus we compute this index by summing the log pointwise predictive density in each dimension and the result is shown in Table \ref{tab:log_predictive_sphere}. 

Moreover, discrete Fr\'{e}chet distance is a metric to measure the similarity between two curves which considers the locations and orderings between predictions and real data. This method is widely used in GPS data, especially for comparing trajectories of animals, vehicles or other moving objects \cite{devogele2017optimized}. The results show that our model provides small predictive error under this metric. In other words, our model is effective even the measurement is different from its objective function. The numerical results are shown in Table \ref{tab:log_predictive_sphere}.

We also show the RMSE of $\cuu{\beta}$ in Table \ref{tab:log_predictive_sphere}. We can see that, when the number of curves is fixed, the RMSE between estimated $\beta$ and real $\beta$ decreases with the increasing number of data points; however, when the number of data points is fixed, the RMSE between estimated $\beta$ and real $\beta$ is almost the same (0.1779 compared to 0.1773 and 0.0212 compared to 0.0214).

\begin{table}[h]
    \centering
    \begin{tabularx}{\textwidth}{Y Y Y Y}
    \hline
          &Type \rom{1} &Type \rom{2} &Type \rom{3} \\
    \hline
    Log predictive likelihood \\
    \hline
    30 curves, 20 points &38.4998 &11.2927 &35.9895 \\
    \hline
    30 curves, 60 points &42.1669 &13.1486 &46.1013 \\
    \hline
    60 curves, 20 points &40.0731 &12.7354 &35.9048 \\
    \hline
    60 curves, 60 points &45.3947 &14.1749 &52.6215 \\
    \hline
    Curve similarity \\ 
    \hline
    30 curves, 20 points &0.0657 &0.0646 &0.0498 \\
    \hline
    30 curves, 60 points &0.0306 &0.0529 &0.0336 \\
    \hline
    60 curves, 20 points &0.0520 &0.0596 &0.0535 \\
    \hline
    60 curves, 60 points &0.0247 &0.0422 &0.0390 \\
    \hline
    RMSE between $\beta$ and $\hat{\beta}$\\
    \hline
    30 curves, 20 points &0.1779 &0.1774 &0.1778 \\
    \hline
    30 curves, 60 points &0.0230 &0.0212 &0.0236 \\
    \hline
    60 curves, 20 points &0.1773 &0.1770 &0.1773 \\
    \hline
    60 curves, 60 points &0.0228 &0.0214 &0.0213 \\
    \end{tabularx}
    \caption{Log-predictive likelihood, discrete Fr\'{e}chet distance between estimated covariance structure and the real covariance structure for several models with three types of data. Root-mean-square-error between estimation and the real value of $\beta$ for several models with three types of data.}
    \label{tab:log_predictive_sphere}
\end{table}

\subsection{Regression Analysis on Kendall's Shape Space}

\begin{figure}[h]
    \centering
    \includegraphics[width=150mm,scale=0.5]{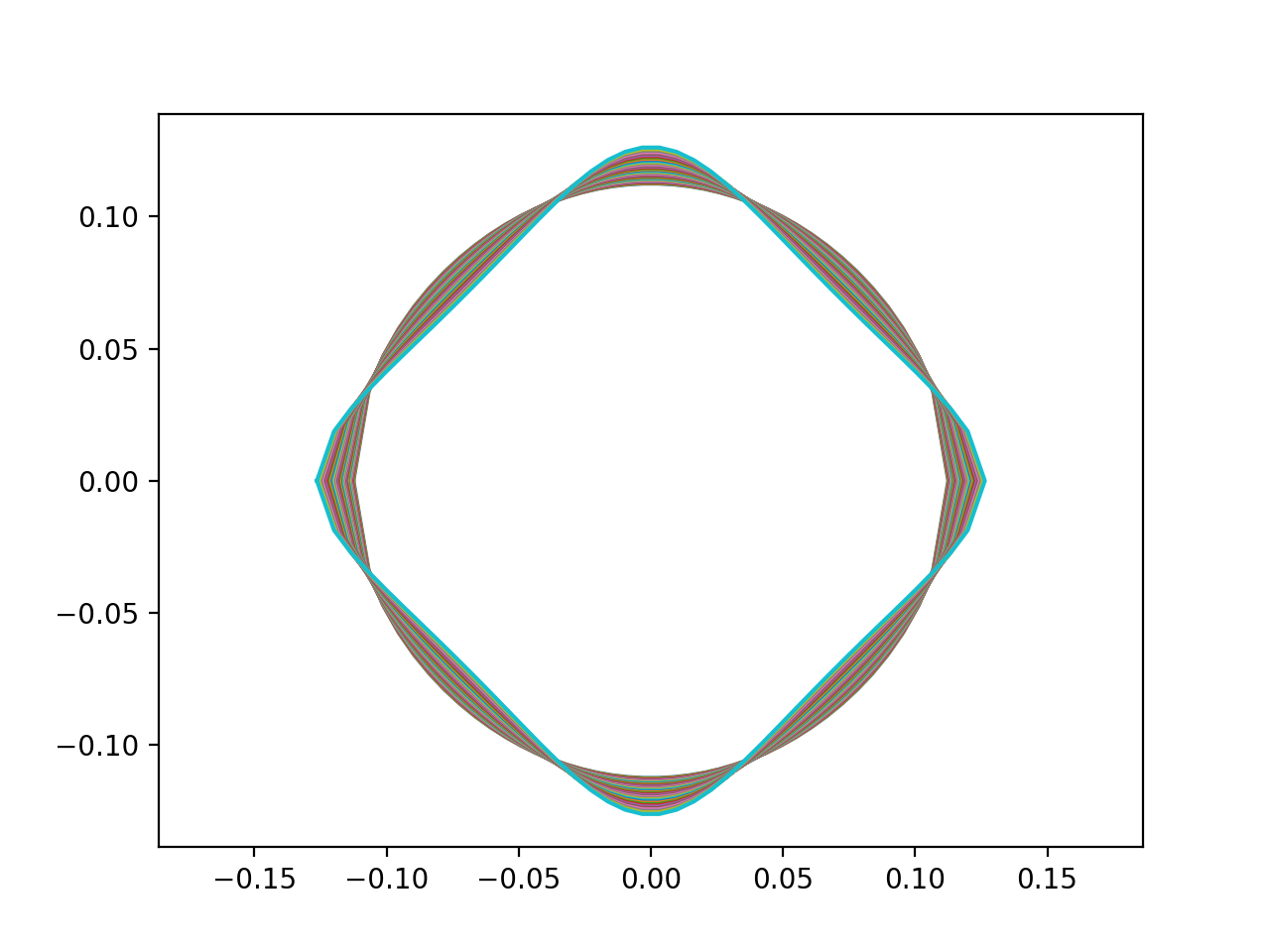}
    \caption{Generated $\mu_0(t)$ on a Kendall's shape space.}
\label{fig:shape_mu_0}
\end{figure}

We firstly generated $\mu_0(t)$, for $t\in (0,1)$, which describes a varying shape. Specifically, the shape is a circle at the beginning and becomes a square in the end. The varying shapes are shown in Figure \ref{fig:shape_mu_0} and they are generated based on an elliptic equation after removing scale, rotation and translation.

We generated data corresponding to 80 landmarks, given by an element of $\mathbb{C}^{2\times 80}$(See the Appendix for more details about Kendall's shape space)
and the mean shape function $\mu_0(t)$ is defined as
\begin{gather}
    \mu_{0,z}(t) =
    \begin{cases}
    (\frac{z}{20}-1,\sqrt{(\frac{z}{20}-1)^2+(1-(\frac{z}{20}-1)^2)(1-\frac{t}{2})^2}i), & \text{for } z=1,...,40 \\
    (1-\frac{z}{20},-\sqrt{(1-\frac{z}{20})^2+(1-(\frac{z}{20}-1)^2)(1-\frac{t}{2})^2}i), & \text{for } z=41,...,80
    \end{cases}
\end{gather}

Then, we use the scalar covariates to generate tangent vectors from the mean function $\mu_0(t)$ to mean structures $\mu_1(t)$ and $\mu_2(t)$ by
\begin{gather}
\begin{split}
    \mu_1(t) &=\Exp(\mu_0(t), \sum_{p=1}^Pu_{1,p}\sin{(t)}^3\sin{(r(t))}), \\
    \mu_2(t) &=\Exp(\mu_0(t), \sum_{p=1}^Pu_{2,p}\sin{(t)}^3\sin{(r(t))}\cos{(r(t))}),
\end{split}
\end{gather}
where $P$ refers to the number of elements in $\mathbf{u}_1$, $u_{1,1}=0$, $u_{1,2}=0$, $u_{2,1}=1$, $u_{2,2}=2$ and $r(t)$ refers to a ranking function of $t$. For example, when $t=t_3$, $r(t)=3$. Thus, we obtain $3$ curves in a batch and each curve is consist of $10$ points (curves). 

\begin{figure}[h]
    \centering
    \includegraphics[width=150mm,scale=0.5]{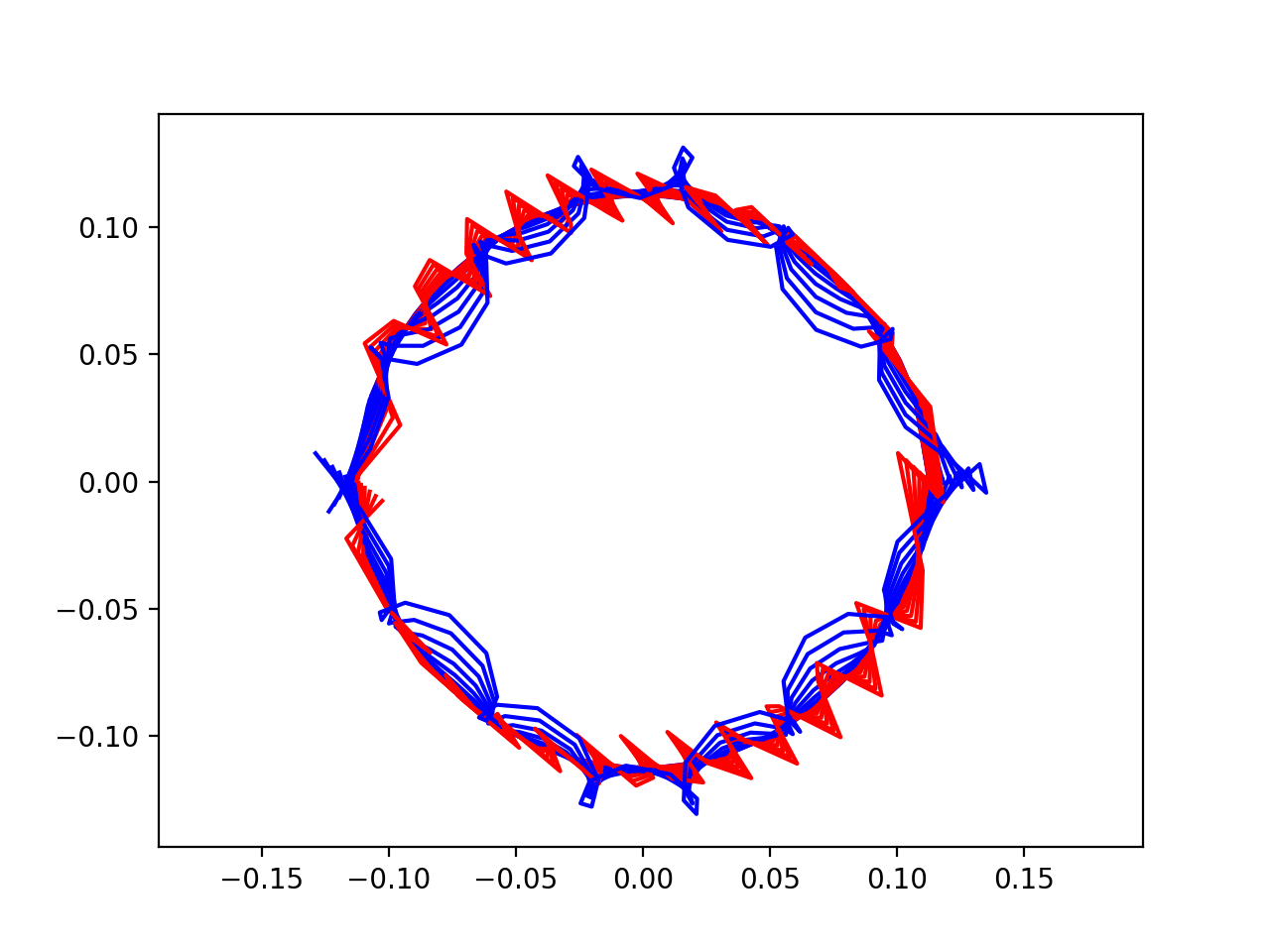}
    \caption{Generated mean structures on Kendall's shape space.}
\label{fig:shape_mu_1_mu_2}
\end{figure}

The generated $\mu_0(t)$ and the mean structures $\mu_1(t)$ and $\mu_2(t)$ are shown in Figures \ref{fig:shape_mu_0} and \ref{fig:shape_mu_1_mu_2} respectively.

We estimate the mean structure and covariance structure as discussed in Section \ref{sec:estimated_mean_structure} and Section \ref{sec:estimate_covariance_structure}. The training data consist of all data in batch one together with the first and the second curve in batch two; the difference between these types of prediction is the handling of the third curve in batch two. Specifically, for Type \rom{1} prediction, $8$ time points are randomly selected to form the training data and the remaining $2$ points are used as the test data; with regard to Type \rom{2} prediction, we add the first half into the training data and the remaining for the test data; as for the Type \rom{3} prediction, the last $3$ time points are considered as the test data and the remaining are added into the training data. Each numerical experiment is repeated for $100$ times. In addition, to compare the performances among different regression models, we also tested the WGPFR, approximate wrapped Gaussian process functional regression model, functional linear regression model on \rma\ (the mean structure) and wrapped Gaussian process Fr\'{e}chet mean regression model. 

The RMSE are given in Table \ref{tab:rmse_r_shape_data}. Similar to the simulation study on a $S^2$, the WGPFR model achieves the best prediction results in both interpolation and extrapolation. Moreover, the time consuming of simulation study on $S^2$ is less than $10$ minutes while that of shape space is about one hour, since the later has higher dimension. Specifically, there are $3$ sets of estimated hyper-parameters for a wrapped Gaussian process on a $S^2$ while $160$ sets of estimated hyper-parameters for a wrapped Gaussian process on a shape space.

\begin{table}[h]
    \centering
    \begin{tabularx}{\textwidth}{Y Y Y Y}
    \hline
           &Type \rom{1}   &Type \rom{2}  &Type \rom{3}  \\
    \hline
    10 curves, 20 points\\
    \hline
    WGPFR   &\textbf{0.0189}  &\textbf{0.0197} &\textbf{0.0202} \\
    FLRM    &0.2702 &0.2777 &0.3425 \\
    WGFmR   &0.1091 &0.2316 &0.2453 \\
    \hline
    10 curves, 40 points\\
    \hline
    WGPFR   &\textbf{0.0159} &\textbf{0.0183} &\textbf{0.0192} \\
    FLRM    &0.2663 &0.1206 &0.1767 \\
    WGFmR   &0.0914 &0.1312 &0.1481 \\
    \hline
    20 curves, 20 points\\
    \hline
    WGPFR   &\textbf{0.0178} &\textbf{0.0190} &\textbf{0.0200} \\
    FLRM    &0.2648 &0.2735 &0.2882 \\
    WGFmR   &0.0855  &0.1698 &0.1531 \\
    \hline
    20 curves, 40 points\\
    \hline
    WGPFR   &\textbf{0.0101} &\textbf{0.0148} &\textbf{0.0163} \\
    FLRM    &0.2036 &0.1165 &0.1486 \\
    WGFmR   & 0.0653 &0.1101 &0.1383 \\
    \hline
    \end{tabularx}
    \caption{Root-mean-squared-error of several models with four types of shape data sets.}
    \label{tab:rmse_r_shape_data}
\end{table}

Thus we also show a log pointwise predictive density and the results are in Table \ref{tab:log_predictive_shape}.

\begin{table}[h]
    \centering
    \begin{tabularx}{\textwidth}{Y Y Y Y}
    \hline
          &Type \rom{1} &Type \rom{2} &Type \rom{3} \\
    \hline
    10 curves, 20 points &80.8193 &23.3038 &79.0855 \\
    \hline
    10 curves, 40 points &81.1700 &25.9431 &83.1195 \\
    \hline
    20 curves, 20 points &75.9353 &22.8560 &78.5120 \\
    \hline
    20 curves, 40 points &77.1970 &24.7091 &82.4506 \\
    \hline
    \end{tabularx}
    \caption{Log-predictive likelihood between estimated covariance structure and the real covariance structure for several models with three types of shape data.}
    \label{tab:log_predictive_shape}
\end{table}

\subsection{Flight Trajectory Data}
\label{sec:real data}

\begin{figure}[h]
    \centering
    \includegraphics[width=150mm,scale=0.5]{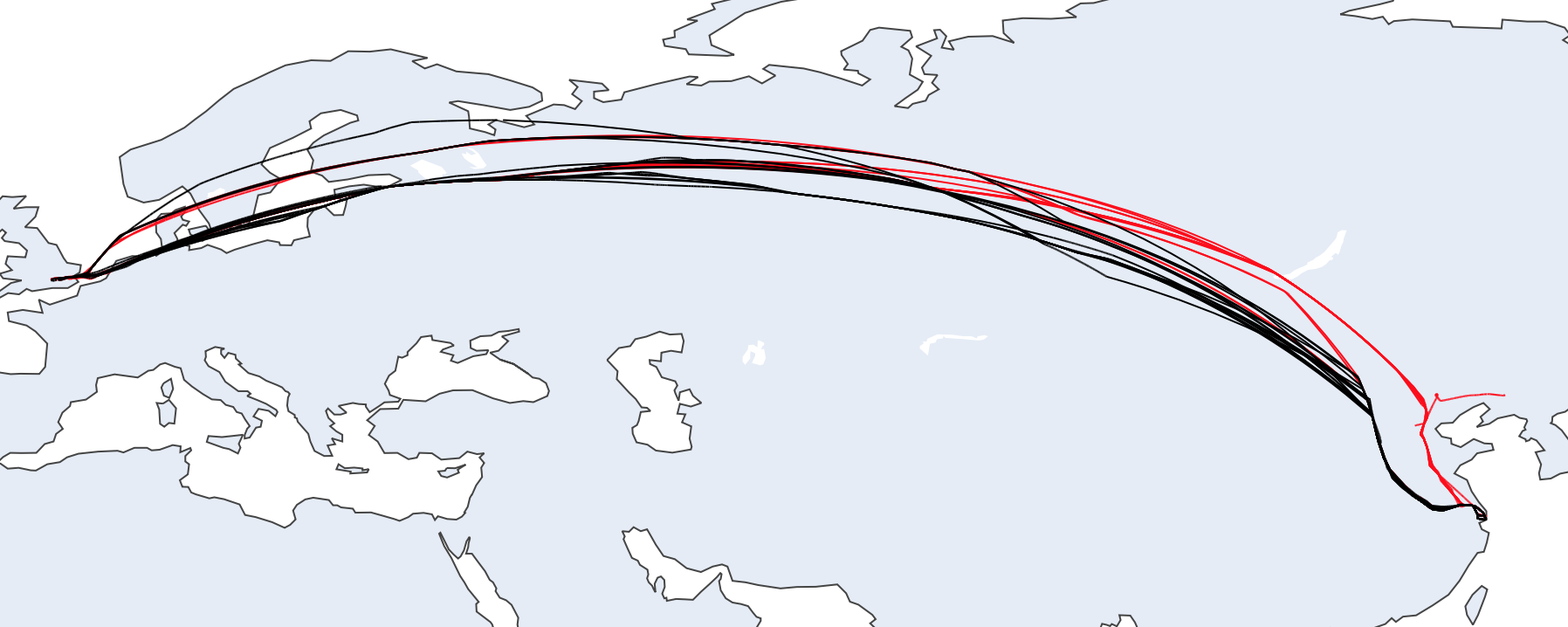}
    \caption{Flight trajectories from Shanghai to London.}
\label{fig:flight_routes}
\end{figure}

In this section, we test our model on a real data set. The earth is roughly a sphere and can be modelled as a copy of $S^2$. Certain data sets, for example hurricane trajectories, are therefore considered as a manifold-valued random curve \cite{su2014statistical}. Here we study flight trajectory data, shown in Figure \ref{fig:flight_routes} in which the red curves represent flights from Shanghai to London by British Airways and the black curves represent flight trajectories of Eastern China Airlines between the same destinations (the data were downloaded from www.variflight.com on 2020). Therefore, these sets of trajectories can naturally be split into two batches and the model with common mean structure can be used. 

The original data includes time, height, speed, longitude and latitude of each flight at discrete time points. We select the position of each airplane as the response variable, which can be transformed onto $S^2$ using longitude and latitude; in addition, the company and time are regarded as non-functional and functional covariates respectively. Before we train the model, it is necessary to pre-process the raw data. In this step, we firstly set the takeoff time of each flight to be $0$ and the landing time to be $1$, excluding taxi time. $25$ trajectories of each company were selected in which the number of observed data points in every flight is greater than $600$. In order to obtain smooth manifold-valued curves from the data, some kernel smoothing functions with small bandwidth were applied to the longitude and latitude of the training data. For computational reasons, we choose every $6$-th data points of each smoothed trajectory as training data ($100$ data points in each curve totally).

To model the mean structure for flight trajectory data, we use company as the batch-specific covariate. In practice, for Eastern China Airlines, the covariate is defined as $0$; and for British Airways, the covariate is defined as $1$. Estimation of the mean structure and covariance structure were described in Section \ref{sec:estimated_mean_structure} and \ref{sec:estimate_covariance_structure}, respectively. The parameters of the basis functions in mean structure and hyper-parameters in the covariance structure were be updated iteratively, as described in Section \ref{sec:update}. In addition, the initial values of hyper-parameters in convariance structure are drawn from a standard Gaussian distribution independently. Afterwards, the predictions of the WGPFR model has been compared to FLRM and WGFmR for the same flight trajectory data.

The overall prediction of WGPFR model should be better than the other models, since the FLRM model only learns from mean structure and ignores the dependent error. In addition, the mean function of WGFmR is a point on $S^2$ and the prediction should have significant error when data are far away from the Fr\'{e}chet mean. This is verified by numerical results in Table \ref{tab:rmse_flight_data}. 

The training data and test data are generated in the following way. We randomly selected another flight trajectory of British Airways which had $>600$ time observations. After the same pre-processing steps, $50$ data points were added into the training data and the remaining $50$ data points are used as the test data. In order to test the performance of interpolation and extrapolation for the real data set, we form the Type \rom{1} prediction by randomly choosing these $50$ data points and form the forecasting by selecting the first $50$ points in the trajectory as training data. In addition, we test the capability for short-term forecasting and long-term forecasting by supposing different test data. Specifically, the $50$-th to $60$-th points and the $50$-th to final data points of the flight trajectory can be selected respectively to form two scenarios of prediction which are denoted as short-term and long-term, respectively.

The predictive accuracy of WGPFR compared to FLR and WGFmR on the flight trajectory data is show in Table \ref{tab:rmse_flight_data}, which shows the mean of  root-mean-squared-error of $20$ repeated simulations to reduce the effects of random seeds. We can see that the WGPFR model outperforms FLR and WGFmR for interpolation. For the short-term prediction, the RMSE is much smaller than that of the long-term prediction. As mentioned previously, the reason is that GPR provides little extra information when the test data are distant from the training data. The prediction of WGFmR is less accurate since the mean structure (Fr\'{e}chet mean) is only a manifold-valued point and the test data set are not close to that point.

\begin{table}[h]
\centering
\begin{tabularx}{\textwidth}{Y Y Y Y}
\hline
           &Type \rom{1} &Short-term &Long-term\\
\hline
    WGPFR  &\textbf{0.0048} &\textbf{0.0096} &\textbf{0.0233} \\ 
    FLRM    &0.0082 &0.0131 &0.0471 \\
    WGFmR  &0.0187 &0.0747 &0.3569 \\
\hline
\end{tabularx}
\caption{Root-mean-squared-error of several models for flight trajectory data.}
\label{tab:rmse_flight_data}
\end{table}

\section{Conclusion and discussion}
\label{sec:discussion}
In this paper we studied a novel functional regression model within a probabilistic framework for nonlinear manifold-valued response variables and real-valued covariates. Specifically, a wrapped Gaussian process functional regression (WGPFR) model is proposed to model the regression relationship into mean structure and covariance structure. For the mean structure, we proposed a functional version of geodesic regression model for the manifold-valued functional response and scalar covariates, where the inference of the mean structure is based on the intrinsic Fr\'{e}chet mean function and the traditional functional linear regression  model is used on the tangent vectors of the manifold-valued data at the intrinsic Fr\'{e}chet mean function to estimate the scalar covariates in an ambient Euclidean space. A wrapped Gaussian process prior is proposed to model the covariance structure in the tangent space at the mean function of each batch, where inference is conducted in a Bayesian way.Furthermore, an iterative approach based on a gradient descent algorithm or variational gradient descent algorithm are also applied to update mean structure and covariance structure efficiently. The mean structure captures the relation between functional dependent variable and batch-based scalar independent variable. Meanwhile, the covariance structure models the nonlinear concurrent relationship between functional output and multiple functional covariates. This idea in WGPFR model avoids the curse of dimensionality by using multiple functional covariates as input for functional regression, which promotes the flexibility of this method. 
 
Future research endeavors could encompass various extensions and enhancements to the proposed WGPFR model. Initially, the WGPFR model, as delineated in this paper, is predicated on the assumption of Riemannian manifolds with infinite injectivity radius. The potential exists to adapt this model to other manifolds that may not conform to this assumption.

Secondly, the current paper presumes the independence of Gaussian processes across different dimensions. Future investigations could explore methodologies as suggested by  \cite{alvarez2012kernels} and \cite{van2020framework}, where Gaussian processes exhibit dependence. However, such an approach would necessitate the development of additional computational efficiencies.

Thirdly, the proposed model could be applied to other intriguing real data sets. For instance, the performance of our model could be evaluated on medical imaging data. Repeated measurements of functional magnetic resonance imaging (fMRI), for example, could be construed as data residing in a specific manifold. Furthermore, the recurrently scanned shapes of the corpus callosum from diverse patients could be modeled to predict the likelihood of Alzheimer’s disease based on certain relative factors \cite{Banerjee_2016_CVPR}. In addition, consideration could be given to non-Gaussian data as \cite{wang2014generalized} and alternative methods to define mean and covariance structures.

The convergence of parameter estimation is substantiated in Theorem \ref{theorem_1} and \ref{theorem_2}. However, the assurance of convergence for the iterative optimization Algorithm \ref{alg:approx_model_rma} remains an open question for future exploration. \cite{hao2018simultaneous} demonstrated that, given suitable initial parameters, the estimation error of conditional maximization problems is confined within the bounds of statistical error and optimization error with a high degree of probability, potentially laying the groundwork for convergence. Nonetheless, it is crucial to acknowledge the distinction between graphical spaces and Riemannian manifolds. The approach proposed by \cite{wang2022gaussian}, which establishes that the convergence rate of GPR is optimal under certain preconditions and is upper bounded based on the smoothness of the correlation function in Euclidean space, could potentially be extrapolated to Riemannian manifolds to facilitate the convergence of covariance structure estimation. This proposition, however, requires rigorous validation.

\section*{Acknowledgments}
JQS’s work is supported by funds of National Key R\&D Program of China (2023YFA1011400), the National Natural Science Foundation of China (No. 12271239) and Shenzhen Fundamental Research Program (No. 20220111). Chao Liu is supported in part by China Postdoctoral Science Foundation, No.2021M691443, No.2021TQ0141 and SUSTC Presidential Postdoctoral Fellowship.


\newpage

\appendix

\section*{Appendix}
\subsection*{Comparison of A Model with or without A Manifold Structure}
In Kendall's shape space, the prediction of a regression model without a manifold structure loses the shape framework while a model with manifold structure still keep it.

\begin{figure}[H]
    
    \centering
    \includegraphics[width=150mm,scale=0.5]{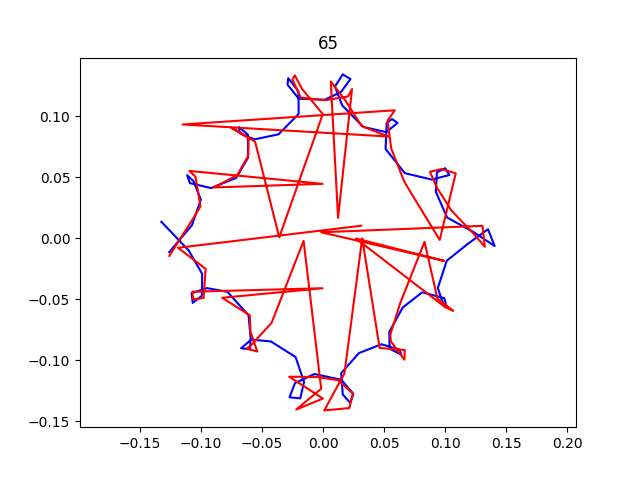}

    \caption{The blue shape is prediction of a Gaussian process regression with a manifold structure while the red shape is prediction of a Gaussian process regression without a manifold structure.}
    
    \label{fig:shape}
\end{figure}

\label{app:theorem}

\subsection*{Examples of Riemannian Manifolds}

\textbf{Sphere $S^n$:}
The two-dimensional sphere $S^2=\{x\in\mathbb{R}^3:\|x\|=1\}$ is a simple example of a Riemannian manifold. 
It is easy to show that the tangent vectors at a point $p\in S^2$ are the vectors which are orthogonal to $p$. 
A Riemannian metric is inherited from the ambient Euclidean metric on $\mathbb{R}^3$, and it exactly the Euclidean inner product between tangent vectors. 
It is easy to show, using this metric, that the shortest curve passing between two points $p,q\in S^2$ is a \xie{the great circle}, which is contained within the intersection of $S^2$ and a plane containing $p$, $q$ and the origin. 
The geodesic is unique exactly when one such plane exists, or in other words, when $p$ and $q$ are not antipodal. 
The formula of geodesic distance is given by
\begin{gather}
	d_\M(p, q)=\cos^{-1}(p^T q). 
\end{gather}
The formula of exponential map is
\begin{equation}
	\Exp(p, v)=\cos(\Vert v\Vert)p + \sin(\Vert v\Vert)\frac{v}{\Vert v\Vert},\  \text{for all\ } v\in T_p\M=\{v:p^Tv=0\}.
	\label{eq:exp_sphere}
\end{equation}
The logarithm map is defined for all pairs $p,q$ which are not antipodal:
\begin{gather}
	\Log(p,q)=\frac{u}{\Vert u\Vert}d_\M(p,q), \ \text{where}\ u=q-(p^Tq)p,\ \text{for all\ }p+q\neq 0.
	\label{eq:log_sphere}
\end{gather}


These concepts can be to the hyper-sphere $S^n$, where $n>2$. 

\textbf{Kendall's Shape Space:}
As a well-developed probability space, Kendall's shape space provides a very useful space for illustrate new theory and applications in statistics and machine learning. 
In 2-dimensional shape analysis \cite{dryden2016statistical}, a landmark is a point that can be unambiguously identified on all objects in a population. \xie{Landmarking} is the process of identifying and labelling a set of $k$ landmarks on each object in an analysis.
This process generates a $2k$ dimensional (real) vector, which can also be represented in a $k$-dimensional complex space $\mathbb{C}^k$.
The \xie{shape} of an object is the equivalence class of the object when translation, rotation and scaling of the object are removed. 
Translation is removed by moving the centroid of each object to be origin point. The landmark vectors then lie in a complex subspace $V=\{(z_i)\in\mathbb{C}^k|\sum_{i=1}^k z_i=0\}$, which is a copy of $\mathbb{C}^{k-1}$. Two configurations in $\mathbb{C}^{k-1}$ are equivalent if they are related by scaling and rotation. The set of equivalence classes can be show to be equal to \xie{complex projective space} $C\mathbb{P}^{k-2}$ and is known as Kendall's shape space \cite{kendall1984shape}. The following formulae of exp map, log map on Kendall's shape space are from \cite{zhang2013probabilistic}.


Analogously to the sphere, we write down formulae which specify the Riemannian geometry on Kendall's shape space. Suppose $p$ and $q$ are two data points in a general $C\mathbb{P}^{k-2}$, the geodesic distance can be computed by
\begin{gather}
    d_\M(p, q) = \arccos \frac{|p^*q|}{\|p\|\|q\|}
\end{gather}
Note that this expression is invariant under multiplication of $p$ and $q$ by arbitrary non-zero complex constants. 
The inverse exponential map is given by
\begin{equation}
	\Log(p,q)=\frac{\theta\cdot(q-\pi_pq)}{\| q-\pi_pq\|},\ \theta=\arccos{\frac{|\langle p,q\rangle|}{\|p\|\|q\|}},\ \pi_pq=\frac{p\cdot\langle p,q\rangle}{ \|p\|^2}
\end{equation}

The exponential map is defined by
\begin{gather}
	\Exp(p,v)=\cos{\theta}\cdot p + \frac{\|p\|\sin{\theta}}{\theta}v,\ \theta=\|\Log(p,q)\|
\end{gather}

The tangent space $T_p\M$ of a point $p$ has the same dimensionality as the Kendall's shape space.
\subsection*{Proofs}

In order to prove the theorems in Section 3, we need the following assumptions for the Riemannian manifold $\M$ and  the manifold-valued data $y(t)$.
\begin{itemize}
	\item (C1) $\mathcal{M}$ is a closed Riemannian submanifold of a Euclidean space $\mathbb{R}^{d_{0}}$, with geodesic distance $d_{\mathcal{M}}$ induced by the Euclidean metric.
	\item 
	(C2) Trajectories $y(t)$ are continuous for $t \in \mathcal{T}$ almost surely.
	\item
	(C3) For all $t \in \mathcal{T}$, the Fr\'echet mean $\mu_{0}(t)$ and sample  Fr\'echet mean $\hat{\mu}_{0}(t)$ exist and are unique  almost surely.
	\item (C4) Almost surely, trajectories $y(t)$ lie in a compact set $S_{t} \subset B_{\mathcal{M}}\left(\mu_{0}(t), r\right)$ for $t \in \mathcal{T}$, where $B_{\mathcal{M}}\left(\mu_{0}(t), r\right) \subset \mathcal{M}$ is an open ball centered at $\mu_{0}(t)$ with radius $r<\inf _{t \in \mathcal{T}} \operatorname{inj}_{\mu_{0}(t)}$.
	\item  (C5) For any $\epsilon>0$,
	$$
	\inf _{t \in \mathcal{T}} \inf _{p: d_{\mathcal{M}}\left(p, \mu_{0}(t)\right)>\epsilon} M(p, t)-M\left(\mu_{0}(t), t\right)>0 .
	$$
\end{itemize}

\subsection*{Proof of Theorem 1}
 
{\bf Proof}. Recall the notations and the multiple functional regression model $V_m(t)=W_m(t)+e_m(t)$, denote the vector of the realizations at  $t_{i}$, $m=1,\cdot,M$, $i=1,\cdots,n$ of $V_m(t),W_m(t)$ and $e_m(t)$ as $\cuu{V}$, $\cuu{W}$ and $\cuu{E}$, respectively. Then we have
\begin{align}\nonumber
 vec(\hat{B})-vec(B)&=	\bigg((\cuu{\Phi}\otimes U)^\top (\cuu{\Phi}\otimes U)\bigg)^{-1}(\cuu{\Phi}\otimes U)^\top vec(\hat{\cu{V}})-vec(B)\\ \nonumber
 &=\bigg((\cuu{\Phi}\otimes U)^\top (\cuu{\Phi}\otimes U)\bigg)^{-1}(\cuu{\Phi}\otimes U)^\top\bigg(vec(\cuu{W}+\cuu{E})+vec(\hat{\cuu{V}}-\cuu{V})\bigg)-vec(B)\\ \nonumber
  &=\bigg((\cuu{\Phi}\otimes U)^\top (\cuu{\Phi}\otimes U)\bigg)^{-1}(\cuu{\Phi}\otimes U)^\top\bigg((\cuu{\Phi}\otimes U)vec(B)+vec(\cuu{E})+vec(\hat{\cuu{V}}-\cuu{V})\bigg)-vec(B)\\ \nonumber
    &=\bigg((\cuu{\Phi}\otimes U)^\top (\cuu{\Phi}\otimes U)\bigg)^{-1}(\cuu{\Phi}\otimes U)^\top vec(\cuu{E})\\ \nonumber
    &+\bigg((\cuu{\Phi}\otimes U)^\top (\cuu{\Phi}\otimes U)\bigg)^{-1}(\cuu{\Phi}\otimes U)^\top vec(\hat{\cuu{V}}-\cuu{V})\\ \nonumber
    &\dot{=}A_1+A_2 
\end{align}
For the first term, as $e_m(t)$ are assumed to be independent to the covariate $\cuu{u}_m$ for $m=1,\cdots,M$, so that $E(A_1)=0$, $Var(A_1)=O_p(1/M)$. Therefore $A_1=o(1)$ as $M\rightarrow \infty$.

For the second term, note that 
$$
\hat{V}_m(t)-V_m(t)=\Log(\hat{\mu}_0(t),y_m(t))-\Log({\mu}_0(t),y_m(t))\leq d_\M(\hat{\mu}_0(t),\mu_0(t))=O_p(1/\sqrt{M})
$$
where the last term has been proofed in \cite{dai2018principal} under the above conditions (C1)-(C5).
So the second term also has $A_2=o(1)$ as  $M\rightarrow \infty$.  \hfill$\blacksquare$
 
 
\subsection*{Proof of Theorem 2}
 
Before proofing the Theorem 2, we first introduce some notation to simplify the proof.
Recall that $\cuu{\tau}_{md}=(\tau_{md}(t_{m1}),\cdots,\tau_{md}(t_{mn}))$, for simplicity, we omit the subscript $md$ and denote the $n$ observation at the $m$-th batch and the $d$-th dimension as $\cu{z}_n=(z_1,\cdots,z_n)\dot{=}(\tau_{md}(t_{m1}),\cdots,\tau_{md}(t_{mn}))$, and denote the corresponding covariates 
$\cuu{X}_n=(\cuu{x}_1,\cdots,\cuu{x}_n)$ with $\cuu{x}_i\in \mathcal{X}$ which are independently drawn from a distribution $\mathcal{U}(\cuu{x})$. Denote the observed data $\mathcal{D}_n=\{(\cuu{x}_i,z_i),i=1,\cdots,n\}$.

Let $\tau_0(\cdot)$ be the true underlying function. Assume that the underlying process $\tau(\cdot) \sim \operatorname{GP}(0, K(\cdot, \cdot ; \boldsymbol{\theta}))$ where all the subscript $md$ are omitted. Denote
$$
\begin{aligned}
	p_{g p}\left(z_{n}\right) &=\int_{\mathcal{F}} p\left(z_{1}, \ldots, z_{n} \mid \tau\left(\boldsymbol{X}_{n}\right)\right) d p_{n}(\tau) \\
 	p_{0}\left(z_{n}\right) &=p\left(z_{1}, \ldots, z_{n} \mid \tau_{0}\left(\boldsymbol{X}_{n}\right)\right)
\end{aligned}
$$
then $p_{g p}\left(z_{n}\right)$ is the Bayesian predictive distribution of $z_{n}$ based on the GPR model. Note that $p_{n}(\tau)$ depends on $n$ since the hyperparameters of $\tau(\cdot)$ is estimated from the data.
 
{\bf Proof}. 	
It suffices to show $$
\frac{1}{n} E_{\boldsymbol{X}_{n}}\left(D\left[p_{0}\left(\boldsymbol{z}_{n}\right),   p_{g p}\left(\boldsymbol{z}_{n}\right)\right]\right) \rightarrow 0 \quad \text { as } n \rightarrow \infty,
$$
Note that
$$
\begin{aligned}
	D\left[p_{0}\left(\boldsymbol{z}_{n}\right),  {p}_{g p}\left(\boldsymbol{z}_{n}\right)\right] &=\int_{\mathcal{Z}^{n}} p_{0}\left(z_{1}, \cdots, z_{n}\right) \log \frac{p_{0}\left(z_{1}, \cdots, z_{n}\right)}{  p_{g p}\left(z_{1}, \cdots, z_{n}\right)} d z_{1} \cdots d z_{n} \\
	&=\int_{\mathcal{Z}^{n}} p_{0}\left(z_{1}, \cdots, z_{n}\right)\left[-\log    p_{g p}\left(z_{1}, \ldots, z_{n}\right)+\log p_{0}\left(z_{1}, \ldots, z_{n}\right)\right] d z_{1} \cdots d z_{n} .
\end{aligned}
$$
It suffices to give an upper bound for the term $-\log   p_{g p}\left(z_{1}, \ldots, z_{n}\right)+\log p_{0}\left(z_{1}, \ldots, z_{n}\right)$.

Let $\mathcal{H}$ be the Reproducing Kernel Hilbert Space (RKHS) associated with the
covariance function $k(\cdot, \cdot ; \boldsymbol{\theta})$, and $\mathcal{H}_{n}$ the span of $\left\{k\left(\cdot, \boldsymbol{x}_{i} ; \boldsymbol{\theta}\right)\right\}$, i.e. $\mathcal{H}_{n}=\{f(\cdot): f(\boldsymbol{x})=$
$\sum_{i=1}^{n} \alpha_{i} k\left(\boldsymbol{x}, \boldsymbol{x}_{i} ; \boldsymbol{\theta}\right)$, for any $\left.\alpha_{i} \in \mathbb{R}\right\}$. 
Assume the true underlying function $\tau_{0} \in \mathcal{H}_{n}$,
then $\tau_{0}(\cdot)$ can be expressed as
$$
\tau_{0}(\cdot)=\sum_{i=1}^{n} \alpha_{i} k\left(\cdot, \boldsymbol{x}_{i} ; \boldsymbol{\theta}\right) \triangleq K(\cdot) \boldsymbol{\alpha}
$$
where $K(\cdot)=\left(k\left(\cdot, \boldsymbol{x}_{1} ; \boldsymbol{\theta}\right), \ldots, k\left(\cdot, \boldsymbol{x}_{n} ; \boldsymbol{\theta}\right)\right)$ and $\boldsymbol{\alpha}=\left(\alpha_{1}, \ldots, \alpha_{n}\right)^{T} .$ By the properties of
$\mathrm{RKHS},\left\|\tau_{0}\right\|_{k}^{2}=\boldsymbol{\alpha}^{T} \boldsymbol{C}_{n n} \boldsymbol{\alpha}$, and $\left(\tau_{0}\left(\boldsymbol{x}_{1}\right), \ldots, \tau_{0}\left(\boldsymbol{x}_{n}\right)\right)^{T}=\boldsymbol{C}_{n n} \boldsymbol{\alpha}$, where $\boldsymbol{C}_{n n}=\left(k\left(\boldsymbol{x}_{i}, \boldsymbol{x}_{j} ; \boldsymbol{\theta}\right)\right)$ is
the covariance matrix over $\boldsymbol{x}_{i}, i=1, \ldots, n$.

Let $P$ and $\bar{P}$ be any two measures on $\mathcal{F}$, then it yields by Fenchel-Legendre duality relationship that, for any functional $g(\cdot)$ on $\mathcal{F}$,
$$
E_{\bar{P}}[g(\tau)] \leq \log E_{P}\left[e^{g(\tau)}\right]+D[\bar{P}, P]
$$
Let $  g(\tau)=\log p\left(z_{1}, \ldots, z_{n} \mid \tau\right)$ for any $z_{1}, \ldots, z_{n}$ in $\mathcal{Z}$ and $\tau \in \mathcal{F}$, let $P$ be the measure induced by $G P\left(0, k\left(\cdot, \cdot ; \hat{\boldsymbol{\theta}}_{n}\right)\right)$, hence its finite dimensional distribution at $z_{1}, \ldots, z_{n}$ is $\tilde{p}\left(z_{1}, \ldots, z_{n}\right)=N\left(0, \hat{\boldsymbol{C}}_{n n}\right)$, and
$$
E_{P}\left[e^{g(\tau)}\right]=E_{P}\left[p\left(z_{1}, \ldots, z_{n} \mid \tau\right)\right]=\int_{\mathcal{F}} p\left(z_{1}, \cdots, z_{n} \mid \tau\right) d p_{n}(\tau)=p_{g p}\left(\boldsymbol{z}_{n}\right)
$$
where $\hat{\boldsymbol{C}}_{n n}$ is defined in the same way as $\boldsymbol{C}_{n n}$ but with $\boldsymbol{\theta}$ being replaced by its estimator
$\hat{\cuu{\theta}}_{n}$.

Let $\bar{P}$ be the posterior distribution of $\tau(\cdot)$ on $\mathcal{F}$ which has a prior distribution $G P(0, k(\cdot, \cdot ; \boldsymbol{\theta}))$
and normal likelihood $\prod_{i=1}^{n} N\left(\hat{z}_{i} ; \tau\left(\boldsymbol{x}_{i}\right), \sigma^{2}\right)$, where
$$
\hat{\cuu{z}}_n \triangleq\left(\begin{array}{c}

	\hat{z}_{1} \\
	\vdots \\
	\hat{z}_{n}
\end{array}\right)=\left(\boldsymbol{C}_{n n}+\sigma^{2} \boldsymbol{I}\right) \boldsymbol{\alpha}
$$
and $\sigma^{2}$ is a constant to be specified. In other words, we assume a model $z=\tau(\boldsymbol{x})+\eta$
with $\eta \sim N\left(0, \sigma^{2}\right)$ and $\tau(\cdot) \sim G P(0, k(\cdot, \cdot ; \boldsymbol{\theta}))$, and $\hat{\boldsymbol{z}}_n$   is a set of observations at $\cu{x}_{1}, \ldots, \cuu{x}_{n} .$ Thus, $\bar{P}(\tau)=p\left(\tau \mid \hat{\cuu{z}}_n, \cuu{X}_{n}\right)$ is a probability measure on $\mathcal{F}$. Therefore,
by Gaussian process regression, the posterior of $\left(\tau_{1}, \ldots, \tau_{n}\right) \triangleq\left(\tau\left(\boldsymbol{x}_{1}\right), \ldots, \tau\left(\boldsymbol{x}_{n}\right)\right)$ is
$$
\begin{aligned}
	\bar{p}\left(\tau_{1}, \cdots, \tau_{n}\right) & \triangleq p\left(\tau_{1}, \cdots, \tau_{n} \mid \hat{\cuu{z}}, \boldsymbol{X}_{n}\right) \\
	&=N\left(\boldsymbol{C}_{n n}\left(\boldsymbol{C}_{n n}+\sigma^{2} \boldsymbol{I}\right)^{-1} \hat{\boldsymbol{z}}, \boldsymbol{C}_{n n}\left(\boldsymbol{C}_{n n}+\sigma^{2} \boldsymbol{I}\right)^{-1} \sigma^{2}\right) \\
	&=N\left(\boldsymbol{C}_{n n} \boldsymbol{\alpha}, \boldsymbol{C}_{n n}\left(\boldsymbol{C}_{n n}+\sigma^{2} \boldsymbol{I}\right)^{-1} \sigma^{2}\right) \\
	&=N\left(\boldsymbol{C}_{n n} \boldsymbol{\alpha}, \boldsymbol{C}_{n n} B^{-1}\right)
\end{aligned}
$$
where $B=\boldsymbol{I}+\sigma^{-2} \boldsymbol{C}_{n n}$.

It follows that

$\begin{aligned} D[\bar{P}, P] &=\int_{\mathcal{F}} \log \left(\frac{d \bar{P}}{d P}\right) d \bar{P} \\ &=\int_{R^{n}} \bar{p}\left(\tau_{1}, \ldots, \tau_{n}\right) \log \frac{\bar{p}\left(\tau_{1}, \ldots, \tau_{n}\right)}{\tilde{p}\left(\tau_{1}, \ldots, \tau_{n}\right)} d \tau_{1} \ldots d \tau_{n} \\ &=\frac{1}{2}\left[\log \left|\hat{C}_{n n}\right|-\log \left|C_{n n}\right|+\log |B|+\operatorname{tr}\left(\hat{\boldsymbol{C}}_{n n}^{-1} \boldsymbol{C}_{n n} B^{-1}\right)+\left(\boldsymbol{C}_{n n} \boldsymbol{\alpha}\right)^{T} \hat{\boldsymbol{C}}_{n n}^{-1}\left(\boldsymbol{C}_{n n} \boldsymbol{\alpha}\right)-n\right] \\ &=\frac{1}{2}\left[-\log \left|\hat{\boldsymbol{C}}_{n n}^{-1} \boldsymbol{C}_{n n}\right|+\log |B|+\operatorname{tr}\left(\hat{\boldsymbol{C}}_{n n}^{-1} \boldsymbol{C}_{n n} B^{-1}\right)+\left\|\tau_{0}\right\|_{k}^{2}\right.\\ & \left.+\boldsymbol{\alpha}^{T} \boldsymbol{C}_{n n}\left(\hat{\boldsymbol{C}}_{n n}^{-1} \boldsymbol{C}_{n n}-\boldsymbol{I}\right) \boldsymbol{\alpha}-n\right] \end{aligned}$

On the other hand,
$$
E_{\bar{P}}[g(\tau)]=E_{\bar{P}}\left[\log p\left(z_{1}, \ldots, z_{n} \mid \tau\right)\right]=\sum_{i=1}^{n} E_{\bar{P}}\left[\log p\left(z_{i} \mid \tau\left(\boldsymbol{x}_{i}\right)\right)\right]
$$
By Taylor's expansion, expanding $\log p\left(z_{i} \mid \tau\left(\boldsymbol{x}_{i}\right)\right)$ to the second order at $\tau_{0}\left(\boldsymbol{x}_{i}\right)$ yields
$$
\begin{aligned}
	\log p\left(z_{i} \mid \tau\left(\boldsymbol{x}_{i}\right)\right)=& \log p\left(z_{i} \mid \tau_{0}\left(\boldsymbol{x}_{i}\right)\right)+\left.\frac{d\left[\log p\left(z_{i} \mid \tau\left(\boldsymbol{x}_{i}\right)\right)\right]}{d \tau\left(\boldsymbol{x}_{i}\right)}\right|_{\tau\left(\boldsymbol{x}_{i}\right)=\tau_{0}\left(\boldsymbol{x}_{i}\right)}\left(\tau\left(\boldsymbol{x}_{i}\right)-\tau_{0}\left(\boldsymbol{x}_{i}\right)\right) \\
	&+\left.\frac{1}{2} \frac{d^{2}\left[\log p\left(z_{i} \mid \tau\left(\boldsymbol{x}_{i}\right)\right)\right]}{\left[d \tau\left(\boldsymbol{x}_{i}\right)\right]^{2}}\right|_{\tau\left(\boldsymbol{x}_{i}\right)=\tilde{\tau}\left(\boldsymbol{x}_{i}\right)}\left(\tau\left(\boldsymbol{x}_{i}\right)-\tau_{0}\left(\boldsymbol{x}_{i}\right)\right)^{2}
\end{aligned}
$$
where $\tilde{\tau}\left(\boldsymbol{x}_{i}\right)=\tau_{0}\left(\boldsymbol{x}_{i}\right)+\lambda\left(\tau\left(\boldsymbol{x}_{i}\right)-\tau_{0}\left(\boldsymbol{x}_{i}\right)\right)$ for some $0 \leq \lambda \leq 1$.
For Gaussian probability density function, 
it follows that
$$
\begin{aligned}
	E_{\bar{P}}\left[\log p\left(z_{i} \mid \tau\left(\boldsymbol{x}_{i}\right)\right)\right]= \log p\left(z_{i} \mid \tau_{0}\left(\boldsymbol{x}_{i}\right)\right)-\frac{\sigma^2}{2}Var[\tau\left(\boldsymbol{x}_{i}\right)]
\end{aligned}
$$
so that
$$
E_{\bar{P}}\left[\log p\left(z_{1}, \ldots, z_{n} \mid \tau\right)\right]=\log p_0\left(z_{1}, \ldots, z_{n} \right)-\frac{\sigma^2}{2}\operatorname{tr}\left(Var[\tau\left(\boldsymbol{X}_n\right)]\right)=\log p_0\left(z_{1}, \ldots, z_{n} \right)-\frac{\sigma^2}{2}\operatorname{tr}\left(\boldsymbol{C}_{n n} B^{-1}\right)
$$
Therefore,

$\begin{aligned} &-\log p_{g p}\left(z_{1}, \ldots, z_{n}\right)+\log p_{0}\left(z_{1}, \ldots, z_{n}\right) \\ \leq &-\log E_{P}\left[e^{g(\tau)}\right]+E_{\bar{P}}[g(\tau)]+\frac{\sigma^2}{2}   \operatorname{tr}\left(\boldsymbol{C}_{n n} B^{-1}\right) \\ \leq & D[\bar{P}, P]+\frac{\sigma^2}{2}  \operatorname{tr}\left(\boldsymbol{C}_{n n} B^{-1}\right) \\=& \frac{1}{2}\left\|\tau_{0}\right\|_{k}^{2}+\frac{1}{2}\left[-\log \left|\hat{\boldsymbol{C}}_{n n}^{-1} \boldsymbol{C}_{n n}\right|+\log |B|+\operatorname{tr}\left(\hat{\boldsymbol{C}}_{n n}^{-1} \boldsymbol{C}_{n n} B^{-1}+\sigma^2 \boldsymbol{C}_{n n} B^{-1}\right)\right.\\ & \left.+\boldsymbol{\alpha}^{T} \boldsymbol{C}_{n n}\left(\hat{\boldsymbol{C}}_{n n}^{-1} \boldsymbol{C}_{n n}-\boldsymbol{I}\right) \boldsymbol{\alpha}-n\right] \end{aligned}$.

Since the covariance function is continuous in $\boldsymbol{\theta}$ and $\hat{\boldsymbol{\theta}}_{n} \rightarrow \boldsymbol{\theta}$ we have $\hat{\boldsymbol{C}}_{n n}^{-1} \boldsymbol{C}_{n n}-\boldsymbol{I} \rightarrow 0$
as $n \rightarrow \infty$. Therefore there exist some positive constants $K$ and $\epsilon$ such that
$$
\begin{gathered}
	-\log \left|\hat{\boldsymbol{C}}_{n n}^{-1} \boldsymbol{C}_{n n}\right|<K\\  \quad \boldsymbol{\alpha}^{T} \boldsymbol{C}_{n n}\left(\hat{\boldsymbol{C}}_{n n}^{-1} \boldsymbol{C}_{n n}-\boldsymbol{I}\right) \boldsymbol{\alpha}<K \\
	\operatorname{tr}\left((\hat{\boldsymbol{C}}_{n n}^{-1} \boldsymbol{C}_{n n}-I) B^{-1}\right)<K
\end{gathered}
$$
Thus
$$
-\log p_{g p}\left(z_{1}, \ldots, z_{n}\right)+\log p_{0}\left(z_{1}, \ldots, z_{n}\right) <\frac{1}{2}\left\|\tau_{0}\right\|_{k}^{2}+\frac{1}{2}\log |B|+\frac{3}{2}K
$$
for any $\tau_{0}(\cdot) \in \mathcal{H}_{n}$.
Taking infimum over $\tau_{0}$ and applying Representer Theorem (see Lemma 2 in \cite{seeger2008information}) we obtain
$$
-\log p_{g p}\left(z_{1}, \ldots, z_{n}\right)+\log p_{0}\left(z_{1}, \ldots, z_{n}\right) \leq \frac{1}{2}\left\|\tau_{0}\right\|_{k}^{2}+\frac{1}{2} \log \left|\boldsymbol{I}+\sigma^{-2} \boldsymbol{C}_{n n}\right|+\frac{3}{2}K
$$
for all $\tau_{0}(\cdot) \in \mathcal{H}$. 

Therefore, we obtain that
$$
\frac{1}{n} E_{\boldsymbol{X}_{n}}\left(D\left[p_{0}\left(\boldsymbol{z}_{n}\right),  p_{g p}\left(\boldsymbol{z}_{n}\right)\right]\right) \leq \frac{1}{2 n}\left\|\tau_{0}\right\|_{k}^{2}+\frac{1}{2 n} E_{\boldsymbol{X}_{n}}\left(\log \left|\boldsymbol{I}+\sigma^{-2} \boldsymbol{C}_{n n}\right|\right)+\frac{3/2K}{n}\rightarrow 0 
$$
as $ n \rightarrow \infty$.  \hfill$\blacksquare$

\vskip 0.2in

\printbibliography

\end{sloppypar}

\end{document}